\newcommand{\Md}{\dot{\cal M}}

\newcommand{\lsim}{{\textstyle{\; \lower 0.7ex\hbox{$<$}\;
  \atop \raise-0.1ex\hbox{$\sim$}}}}
\newcommand{\gsim}{{\textstyle{\; \lower 0.7ex\hbox{$>$}\;
  \atop \raise-0.1ex\hbox{$\sim$}}}}

\documentclass[onecolumn]{aa}
\usepackage{graphicx}
\usepackage{txfonts}
\begin{document}
\twocolumn
   \title{On the nature of the X-ray source in GK Per}
   \authorrunning{Vrielmann, Ness \& Schmitt}
   \titlerunning{}
%   \subtitle{}

   \author{Sonja Vrielmann\inst{1}
          \and
          Jan-Uwe Ness\inst{2}
          \and
          J\"urgen H.~M.~M. Schmitt\inst{1}
%\inst{2}\fnmsep\thanks{}
          }

   \offprints{S. Vrielmann}

   \institute{Hamburger Sternwarte, Universit\"at Hamburg,
        Gojenbergsweg 112, 21029 Hamburg, Germany\\
        \email{svrielmann@hs.uni-hamburg.de, jschmitt@hs.uni-hamburg.de}
        \and
        Department of Physics, Rudolf Peierls Centre for Theoretical Physics,
        University of Oxford, 1\,Keble Road, Oxford OX1\,3NP, UK
        \email{j.ness1@physics.ox.ac.uk}
             }

   \date{Received; accepted}

   \abstract{We report {\em XMM-Newton} observations of the
   intermediate polar (IP) GK~Per on the rise to the 2002 outburst and
   compare them to {\em Chandra} observations during quiescence.
   The asymmetric spin light curve implies an asymmetric shape of
   a semi-transparent accretion curtain and we propose a model for its
   shape. A low Fe\,{\sc xvii} ($\lambda\lambda$15.01/15.26\,\AA)
   line flux ratio confirms the need for an asymmetric geometry and
   significant effects of resonant line scattering.

   Medium resolution {\em PN} spectra in outburst and {\em ACIS-S}
   spectra in quiescence can both be fitted with a leaky absorber
   model for the post shock hard X-ray emission, a black body
   (outburst) for the thermalized X-ray emission from the white dwarf
   and an optically thin spectrum. The difference in the leaky
   absorber emission between high and low spin as well as
   quasi-periodic oscillation (QPO) or flares states can be fully
   explained by a variation in the absorbing column density.  For the
   explanation of the difference between outburst and quiescence a
   combination of the variation of the column density and the electron
   and ion densities is necessary. The Fe fluorescence at 6.4\,keV
   with an equivalent width of 447\,eV and a possible Compton
   scattering contribution in the red wing of the line is not
   significantly variable during spin cycle or on QPO periods, i.e.\ a
   significant portion of the line originates in the wide accretion
   curtains.

   High-resolution RGS spectra reveal a number of emission lines from
   H-like and He-like elements. The lines are broader than the
   instrumental response with a roughly constant velocity dispersion
   for different lines, indicating identical origin. He-like emission
   lines are used to give values for the electron densities of
   $\log n_e \sim 12$. We do not detect any variation in the emission
   lines during the spin cycle, implying that the lines are not
   noticeably obscured or absorbed. We conclude that they originate in
   the accretion curtains and that accretion might take place from all
   azimuths.

   }

   \maketitle
%
%________________________________________________________________
\section{Introduction}

Cataclysmic variables (CVs) are interacting binaries consisting of a
white dwarf that is accreting matter from a red dwarf star (often a
main sequence star) due to Roche-lobe overflow (for an overview see
Warner 1995).  Intermediate polars are a subgroup of CVs, which
typically have a non-synchronously spinning, magnetised white dwarf,
and - if present at all - a truncated accretion disc. Thus, they fill
the gap between the polars and non-magnetic dwarf novae. Accretion
processes close to the white dwarf surface usually make them strong
X-ray emitters.

%nova, shell
GK Per turned into a nova in 1901 (Hale 1901, Pickering 1901) and is now
observable as an intermediate polar (Watson et al.~1985). However, its
nova shell is still visible at radio, optical, UV and X-ray
wavelengths (Evans et al. 1992, Balman \& \"Ogelman 1999, Anupama \&
Kantharia 2005).  From a study of these shell ejecta Warner (1976)
determined a distance to GK Per of 460\,pc, confirming McLaughlin's
(1960) distance of 470\,pc; only Duerbeck (1981) quotes an unpublished
distance by McLaughlin of 525\,pc.  With an orbital period of about 2
days GK Per lies at the upper end in the period distribution of all
known CVs. Thus, the geometrical dimensions are extremely large for a
CV, noticeable mainly through the fact that the secondary has evolved to a
K1 subgiant (e.g.\ Warner~1976, Morales Rueda et al.~2002).

% Outbursts
GK Per shows dwarf nova type outbursts with a frequency of one in
about 3 years, a typical duration of approximately 50 days (\v{S}imon
2002) and an increase of 2.5 magnitudes in the optical. The slow rise
to maximum light is probably caused by an inside-out outburst (Nogami
et al. 2002) -- in contrast to a more usual outside-in outburst -- or a
more complicated scenario, where the outburst starts at different
distances from the disc centre (\v{S}imon 2002).

%Geometry
Morales Rueda et al. (2002) investigated the absorption spectrum of
GK~Per's companion star, deducing a mass ratio of $q = M_K/M_{wd} =
0.55\pm0.21$ and lower limits for the masses of the stars of $M_K \ge
0.48 \pm 0.32$~M$_\odot$ and $M_{wd} \ge 0.87 \pm 0.24$~M$_\odot$.
The inclination angle is still uncertain, however, it must be smaller
than 73$^\circ$, since no eclipses are observed, and is probably
larger than 50$^\circ$ as can be deduced from models for GK~Per's
accretion geometry (Hellier et al. 2004).

%Spin
During the 1983 outburst Watson et al. (1985) discovered a clearly
periodic signal attributed to the white dwarf spin.  In quiescence,
the spin light curve is double peaked (Patterson 1991, Ishida et
al. 1992), while the outburst spin light curve typically shows a
nearly sinusoidal behaviour (Hellier et al. 2004). Furthermore, during
outburst the full amplitude of the variations is about 50\% of the
maximum value, while during quiescence it drops down to 20\%.

%QPO
In addition to the spin period, quasi-periodic oscillations (QPOs)
have been discovered during outburst in the emission lines by Morales
Rueda et al. (1996). Previously, Watson et al. (1985) reported a
modulation of their EXOSAT (1.5-8.5 keV) data on a time scale of 2000
to $10\,000$\,s.  Subsequently, these QPOs have been observed also in
the optical continuum (Morales Rueda et al. 1999, Nogami et al. 2002)
and in the X-ray band by ASCA (0.7-10~keV, Ishida et al. 1996).

The QPO period varies between 4000 and 6000s, depending on the
spectral range and on the time during the outburst. Hellier et
al. (2004) suggest that the QPOs are caused by bulges travelling with
an orbital period of about 5000s at the inner disc radius (instead of
a bulge rotating with a period of 320 or 380s as suggested by Morales
Rueda et al. 1999). Warner \& Woudt (2002) already suggested such
travelling waves that reach some height above the accretion disc
allowing a temporary blockage of the view towards the white dwarf.

%X-ray spectrum
A few attempts have been made to fit the X-ray continuum spectrum
$\gsim 1$\,keV of GK~Per (EXOSAT, Watson et al. 1985; Norton et
al. 1988; GINGA, Ishida et al. 1992) using power-law and bremsstrahlung
components. The most convincing model appears to be that applied by
Ishida et al. (1992) who fitted a leaky absorber model to the spectral
emission in the range 2-30 keV.
%Fluorescence
Even in early X-ray spectra and all subsequent observations a
prominent Fe fluorescence at 6.4 keV has been detected (e.g.\ Watson
et al 1985; Ishida et al. 1992; Hellier \& Mukai 2004). Hellier \&
Mukai claim that the red wing of the line in their {\em Chandra} data
is due to the movement of the in-falling material.

%UV spectrum
Wu et al. (1989) and Yi \& Kenyon (1997) find a very flat UV spectrum
($\log$ Flux (erg\,cm$^{-2}$\,s$^{-1}$\,\AA$^{-1}$) $\approx$ -11.75
over the range 1000 to 3500\,\AA) which they explain with a
truncated accretion disc. Since a magnetic model is not successful,
they suggest X-ray heating. However, as the increase in X-ray emission
during outburst is insufficient, they suggest that most of the accretion
energy is radiated at extreme UV wavelengths. Yi et al.~(1992) further
find that the observed X-ray flux varies only moderately in spite of
large variations in the mass accretion rate $\Md$ during an
outburst. They explain this observation by obscuration of the hard
X-rays during outburst due to an increased disc thickness.

%this paper
In this paper we re-examine the X-ray characteristics by analysing the
{\it XMM-Newton} EPIC light curves and continuum spectra and
investigate high-resolution RGS line spectra. Mauche (2004) mentioned
these data briefly and confirm the strong Fe K-shell emission at
6.4\,keV. Furthermore, they detect H- and He- like emission lines of
N, O, Ne, Mg, Si and S and possibly He-like Al. Furthermore, we
compare these outburst data to X-ray data during quiescence obtained
with {\em Chandra} ACIS-S. Balman (2001) has analysed in
particular the shell emission and found it to be a multi-temperature
adiabatic shock region with enhanced abundances of Ne and N.

\section{The observations}

%__________________________________________________ One column table
   \begin{table}
      \caption[]{Observations log. All observations were taken on 9
      March 2002. Gaps are not listed, but are obvious from
      Fig.~\ref{Figlcv}.
      }
         \label{Tabobslog}

         \begin{tabular}{lllll}
            \noalign{\smallskip}
            \hline
            \noalign{\smallskip}
            Instrument & Data Mode &
            Filter & Start UT & Stop UT\\
            \noalign{\smallskip}
            \hline
            \noalign{\smallskip}
 M1  &    Imaging          &   Medium C &  14:49:13 & 23:38:53 \\
 M2  &    Imaging          &   Medium C &  14:49:08 & 23:38:53 \\\hline

 OM  &    Imaging \& Fast  &   U        &  14:46:44 & 18:01:00 \\
 OM  &    Imaging \& Fast  &   B        &  18:06:36 & 20:50:52 \\
 OM  &    Imaging \& Fast  &   UVW1     &  20:56:30 & 23:06:46 \\\hline

 PN  &    Imaging          &   Medium   &  15:22:30 & 23:34:10 \\\hline

 R1  &    Spectroscopy     &            &  14:42:54 & 23:42:26 \\
 R2  &    Spectroscopy     &            &  14:42:54 & 23:42:26 \\

            \noalign{\smallskip}
            \hline
         \end{tabular}

   \end{table}
%__________________________________________________ 

GK Per was observed with {\em XMM-Newton} on 9 March 2002 during the
early phase of the 2002 outburst, approximately 1.5 optical magnitudes
below and 30 days before the maximum light. According to the AAVSO
light curves the whole outburst lasted about 70 days.  The data
consist of 32~ks (9~$h$, i.e.\ only about 20\% of a GK Per's orbit)
observing time using all instruments on board XMM-Newton (EPIC PN and
MOS1+2, RGS1+2, OM). The OM data were taken consecutively in the
filters U (7.5~ks), B (9.9~ks), UVW1 (7.8~ks).  For the reduction we
used the standard {\em SAS} software (version 6.0.0). Only for the OM
data we used the pipeline products.

We checked the data for pile-up and found evidence for pile up
for energies above $\sim7$ keV. Thus, during extraction of the spectra
we dismissed central pixels of the stellar image on the CCD chip until
no sign of pile-up was present. Furthermore, at the beginning of the
observing run and between 2 and 5 hours after the start of data taking
the background on the PN CCD chip is unusually high, possibly due to a
minor solar flare. To avoid a contamination of the object spectrum we
only used times with a low background for the extraction of the
spectra.

For the extraction of the light curves, the pile-up and high
background is not a problem, as the flux is conserved
and the light curves are background subtracted.
However, the PN detector suffered failures on short timescales during
the exposure run (at the times of the high background), leading to
drops in the light curve as noticeable in Fig.~\ref{Figlcv} between 0
and 0.5$h$ and between 2 and 5$h$. We still used the full data set for
the light curve extraction. Due to technical problems with the
B filter (it has the highest count rate which lead to artifacts that
could not satisfactorily be corrected) we did not use the B data for
our analysis. No such problems affected the U and UVW1 data.
% PI: Fred Jansen

%During the same outburst, GK Per was observed with {\em CHANDRA} HETG just
%before and after the optical maximum on 27 March 2002 (32.1 ks) and
%9 April 2002 (34.46 ks). \dots --> TBD: Are we going to analyse these data or not?
% PI: Chris Mauche

\begin{figure}
   \centering
  \resizebox{8cm}{!}{\includegraphics{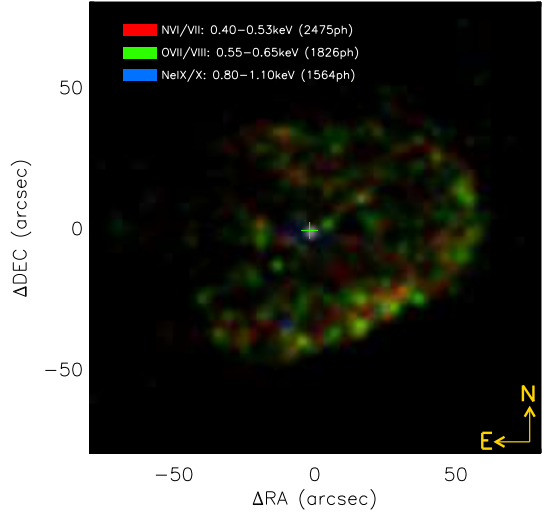}}
      \caption{Smoothed ACIS-S image of GK Per. The colours represent
        different photon energies ranges as indicated in the plot.
%        See Balman (2005) for a more detailed study of the shell spectrum.
              }
         \label{Figshell}
   \end{figure}

A high-resolution image of GK Per was taken with the ACIS-S detector
aboard {\em Chandra} on 10 Feb 2000 for 96.55 ks. Apart from the point
source the expanding shell can clearly be recognised in X-rays. We
colour-coded the photon energies and produced a smoothed image
(displayed in Fig.~\ref{Figshell}), demonstrating that the expanding
shell is a much softer X-ray source than the central star. Contrary to
Balman (2001) we calculated individual spectra for the nova and the
shell by selecting the central source and those parts of the shell
which are brightest in this image (banana shaped region in the south
western quadrant outside a radius of 20 arcsec to avoid
contamination/pile-up of the {\em shell} by the source), respectively.

In order to avoid pile-up of the quiescent {\em source} spectrum we chose an
annular extraction region with an inner radius of 3 pixels ($\approx
1.5$\arcsec), thus excluding the innermost photons from the analysis
which are suffering from pile-up. The effective areas calculated by
the CIAO software account for the fact that the innermost region is
excluded implying the known behaviour of the point spread function and
spectral analyses can still be carried out with lower signal-to-noise.
For a more detailed study of the pile-up problem in this particular
data set see Balman (2005).
% PI: Sölen Balman

%In spite of close examination we could not identify the shell in the
%outburst data, as the contrast between the bright central object to
%the shell is to high. This means that the shell must be fainter than
%...
In order to check whether the emission in the shell was caused by a
reflection effect from a previous outburst, we checked the long term
light curve in the AAVSO Data base. However, the previous outburst
took place nearly one year (March/April 1999) before the {\em Chandra}
quiescent data were obtained (Feb 2000). With a maximal size of 50''
and a distance of 460~pc the light travel time from the nova to the
brighter parts of the shell would be maximally about 130 days which
excludes any re-brightening effects.

%-----------------------------------------------------------
%
%______________________________________________________________

\section{Light curves}

%-----------------------------------------------------------
   \begin{figure*} \centering
   \resizebox{16cm}{!}{\rotatebox{270}{\includegraphics{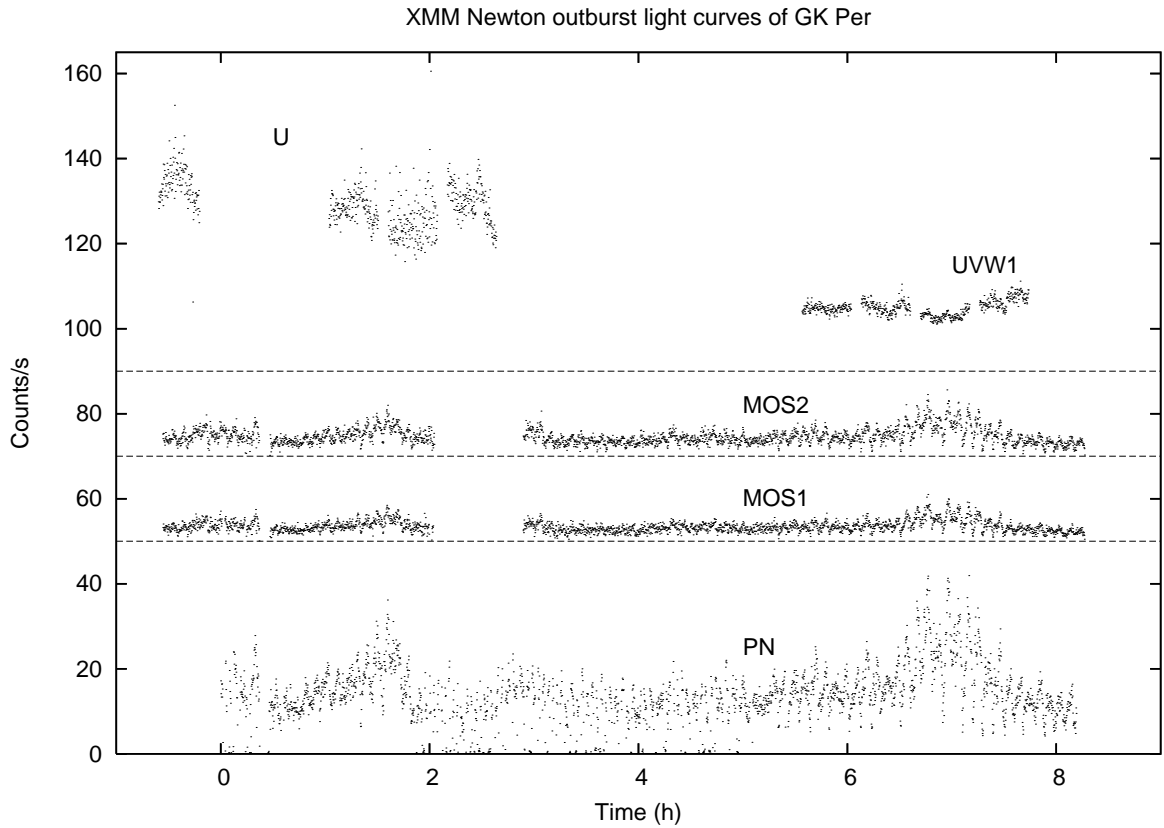}}}

   \caption{EPIC (PN, MOS1 and MOS2) and Optical Monitor (U, UVW1)
              light curves as indicated. All except the PN light
              curves are shifted upwards for clarity of the plot, the
              dashed lines indicate the zero level for the light curve
              immediately above (identical zero line for U and UVW1
              light curve).}  \label{Figlcv} \end{figure*}
%
%______________________________________________________________

In Fig.~\ref{Figlcv} we show the {\em XMM-Newton} X-ray and UV light
curves of GK~Per; $t=0$ corresponds to the start time of the PN observation.
The OM monitor observations consist of 4 stretches of data in
each of the filters U, (B left out) and UVW1. The gaps are actual gaps
in the data acquisition.

%%%FLARES

\begin{figure}
   \centering
   \resizebox{8cm}{!}{\rotatebox{270}{\includegraphics{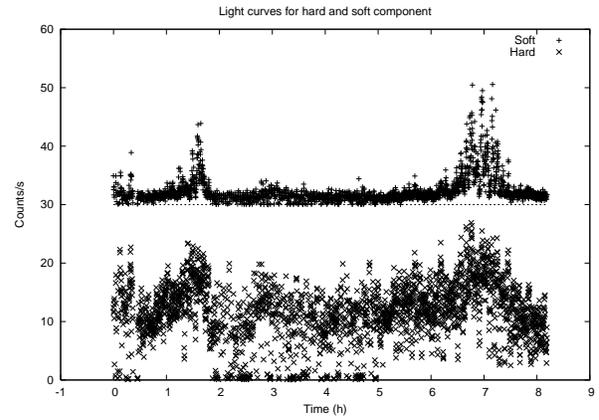}}}

      \caption{Light curve of the soft ($< 1$\,keV, black body,
              shifted upwards by 30 counts/s for clarity) and hard
              ($>1$\,keV, bremsstrahlung) components.  Notice that the
              flares are dominant in the soft component.  }
              \label{Figlcvhardsoft} \end{figure}

The PN light curve shows two prominent peaks - or flares - occurring
at 1.5$h$ and 7$h$, which can also be identified in the MOS light
curves. Note that the background at these times is particularly low,
thus the flares must originate in the source. These X-ray flares
appear to coincide with minima in the UV light curves and might be
linked to the QPO cycle (see below).  To check the
spectral behaviour of the X-ray emission we split up the light curves
into a hard and a soft X-ray component (soft: $< 1$\,keV, hard: $>1$\,keV;
Fig.~\ref{Figlcvhardsoft}).  Clearly, the flares are dominant in the
soft component, i.e.\ the flares must originate mainly from the
thermalized emission on the white dwarf surface.

%-----------------------------------------------------------
%
%______________________________________________________________

%-----------------------------------------------------------
   \begin{figure*}
   \centering
  \resizebox{8cm}{!}{\rotatebox{270}{\includegraphics{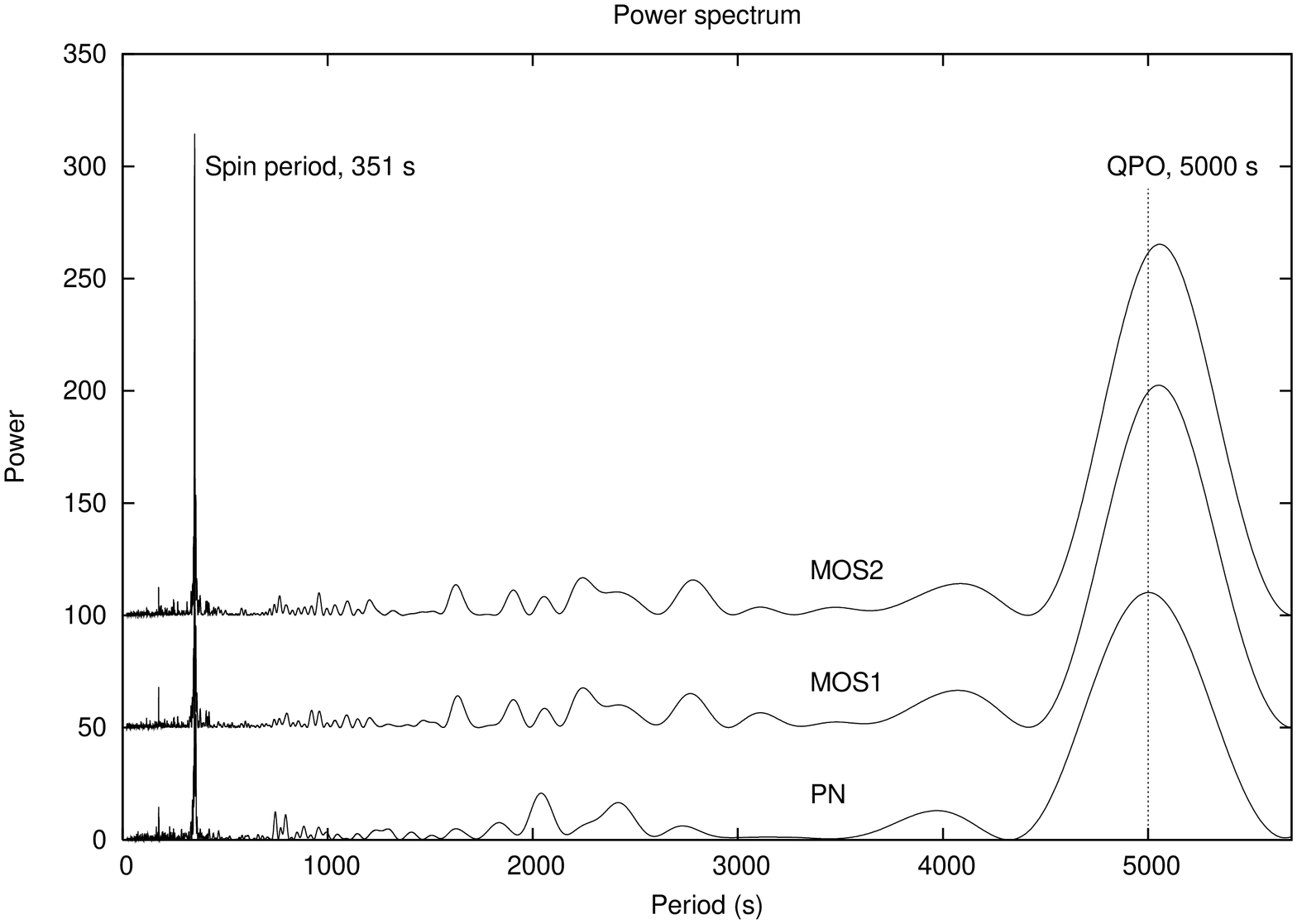}}}
  \resizebox{8cm}{!}{\rotatebox{270}{\includegraphics{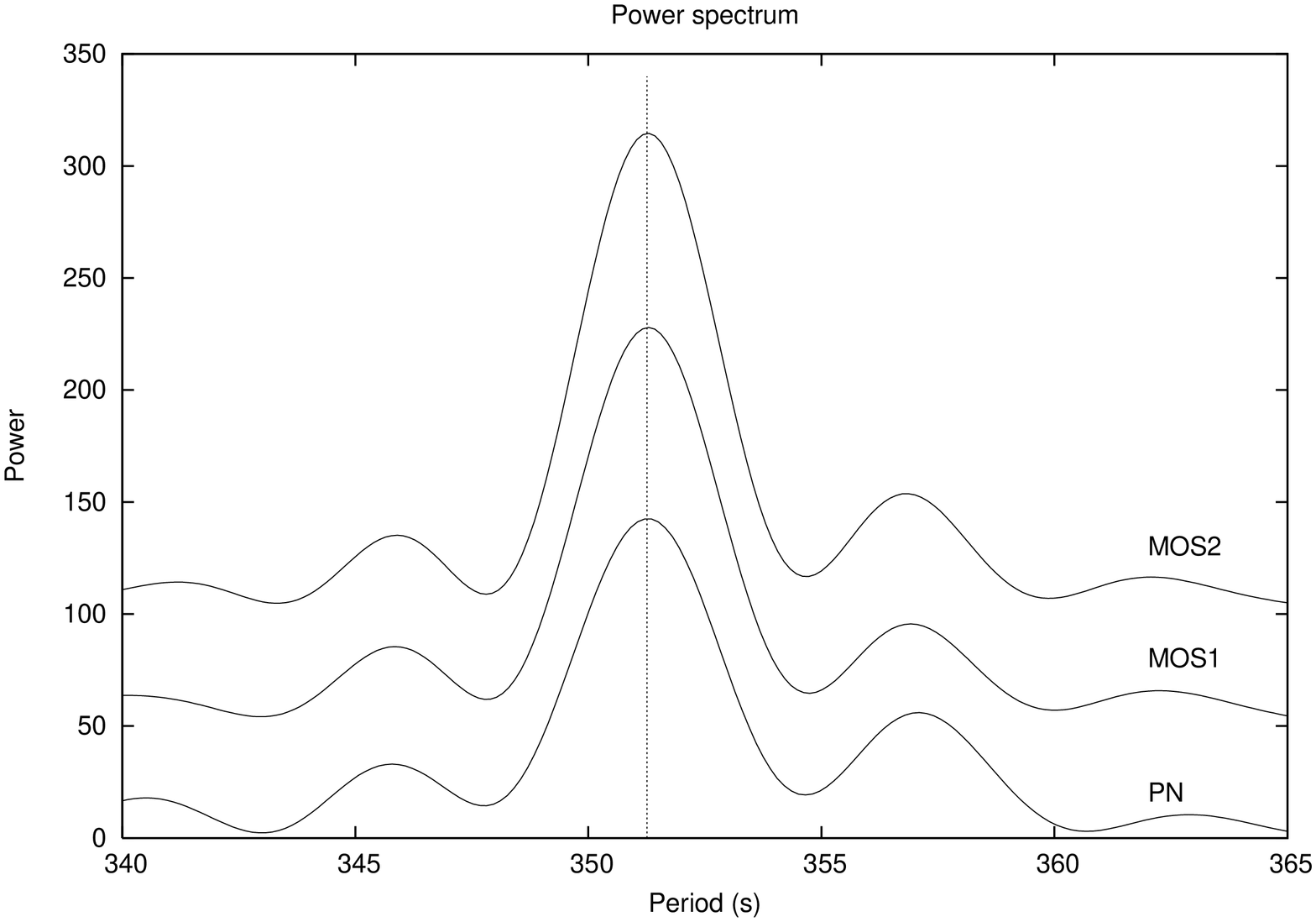}}}

      \caption{Power spectrum of the EPIC data. Clearly visible are
              the fundamental and the first harmonic of the spin
              period at 351.3s and 175.5s. The power spectra for the
              MOS light curve are shifted upwards by 50 units each
              for clarity of the plot. } \label{Figpower} \end{figure*}
%
%______________________________________________________________

%%% SPIN

Consider next the power spectrum of the PN light curve (shown in
Fig.~\ref{Figpower}).  The white dwarf spin period of 351\,s is
clearly visible, the wide width of the power spectrum peak is due to
the short observing run and excludes further analyses of the evolution
of the spin up (Mauche 2004). It is interesting, that the OM data do
not show any significant peak at the spin period of the white dwarf.

%-----------------------------------------------------------
   \begin{figure}
   \centering
  \resizebox{8cm}{!}{\rotatebox{270}{\includegraphics{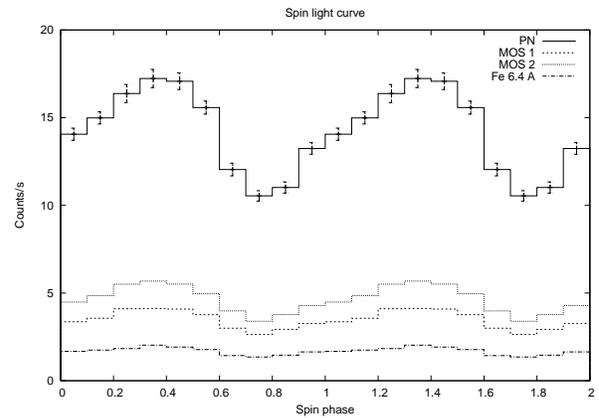}}}

      \caption{EPIC light curve folded on the spin period of the white
              dwarf, 351.26\,s. For clarity, two spin cycles are plotted.
              }
         \label{Figspinlcv}
   \end{figure}
%
%______________________________________________________________

%----------------------------------------------------------- 
   \begin{figure}
   \centering
   \includegraphics[width=8cm]{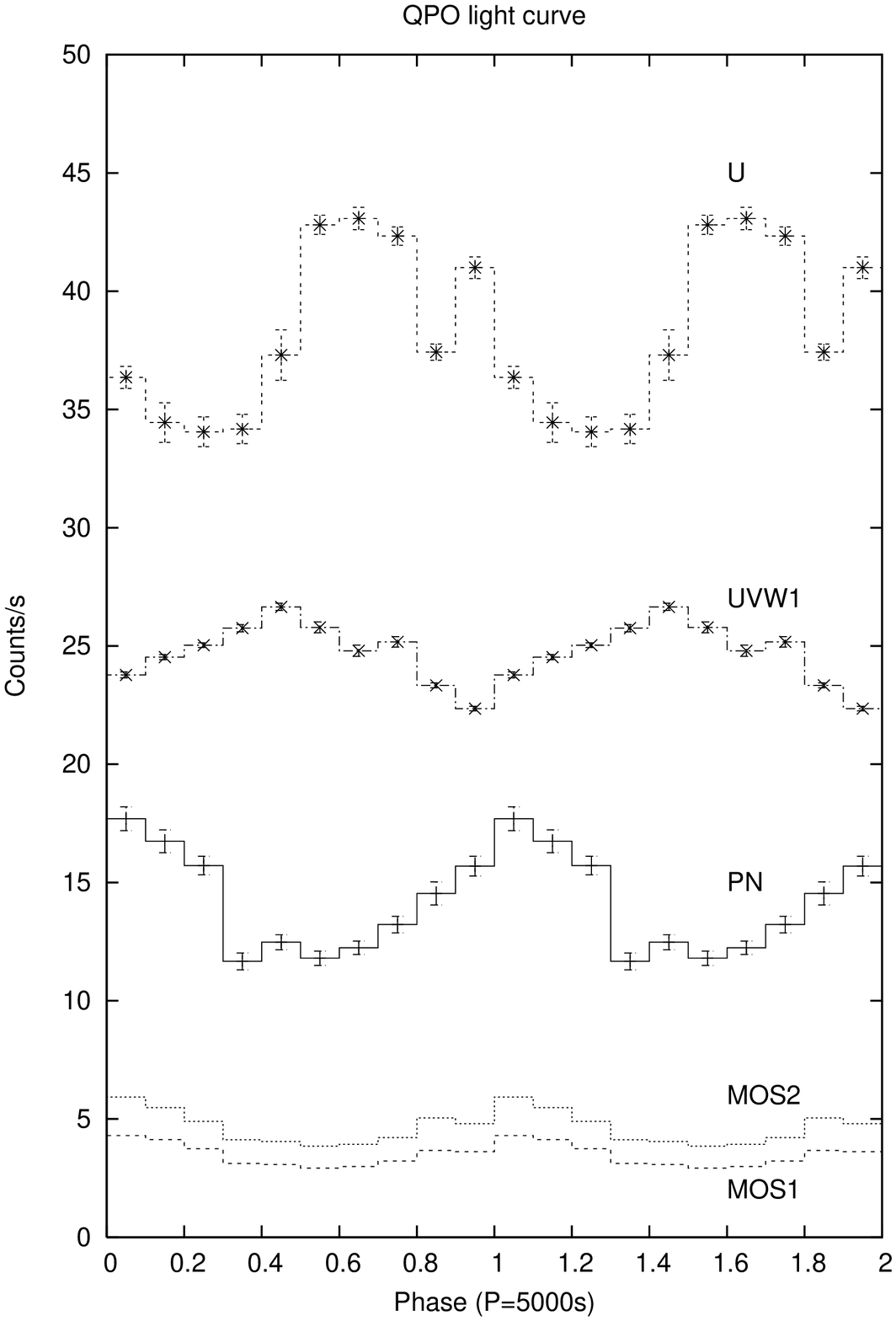}

      \caption{EPIC and OM light curves folded on the 5000~s (QPO) period.
              Note that only the UVW1 light curve is shifted
              upwards by 10 counts/s, all other light curves are at
              their original values. Note also that the UV light
              curves only cover about one QPO cycle each. For clarity,
              two cycles are plotted.
              }
         \label{Figqpolcv}
   \end{figure}
%
%______________________________________________________________

% spin, epic folded on spin period
Fig.~\ref{Figspinlcv} shows the EPIC light curves folded on the spin
period. The full amplitude in the PN light curve is about 39\% of the
maximum value, slightly less than observed by Hellier et al.~(2004)
who report a variation of 50\% at a comparable time during rise into
outburst.  During quiescence, the same variations lie between 15\%
and 20\% (Norton et al. 1988).
% asymmetry of spin profile
Our observed spin profile clearly appears asymmetric: the rise to the
spin light curve maximum is significantly longer than the sudden drop
from maximum to minimum. This shape is obvious in all three EPIC light
curves. While Hellier et al. still find a nearly sinusoidal shape,
their spin light curves show a similar tendency. Also the quiescent
X-ray spin light curve shows the slow rise and sudden drop (Norton et
al. 1988).

%-----------------------------------------------------------
   \begin{figure}
   \centering
  \resizebox{8cm}{!}{\rotatebox{270}{\includegraphics{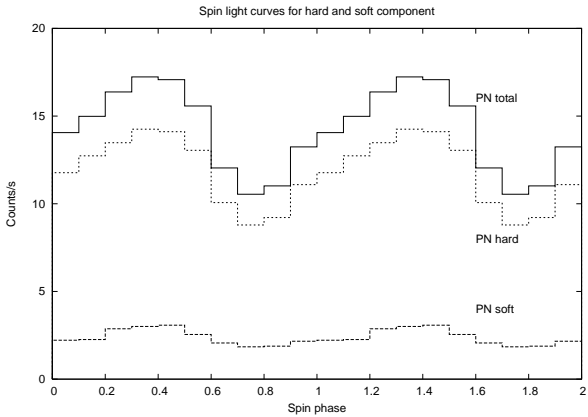}}}

      \caption{Light curves of the soft ($< 1$\,keV) and hard
              ($>1$\,keV) component folded on the Spin period of the
              white dwarf, 351.26\,s. For clarity, two spin cycles are
              plotted.  } \label{Figspinhardsoft} \end{figure}
%
%______________________________________________________________

In order to check for any difference between the soft and hard
components (see next Section), we calculated the light curve for the
%spectral range $< 1$\,keV (soft) and $> 1$\,keV (hard).  As
soft and hard spectral range.  As
Fig.~\ref{Figspinhardsoft} implies, the relative difference between
the two components in terms of the variation of the spin period is
nearly identical and thus identical to the amplitude of the total
variation of 39\%.

%-----------------------------------------------------------
   \begin{figure}
   \centering
  \resizebox{8cm}{!}{\rotatebox{270}{\includegraphics{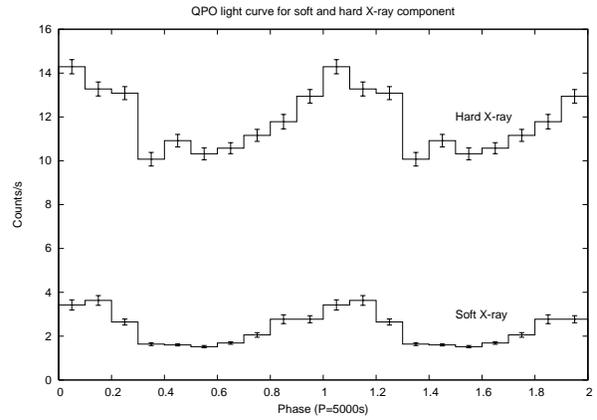}}}

      \caption{Light curves of the soft ($< 1$\,keV) and hard
              ($>1$\,keV) component folded on
              the 5000~s (QPO) period. For clarity, two cycles
              are plotted.
              }
         \label{Figqpohardsoft}
   \end{figure}
%
%______________________________________________________________

%%%  QPOs

Our {\it XMM-Newton} data are compatible with the 5000\,s QPOs
found before (e.g. Morales Rueda et al. 1999, Fig.\ref{Figpower}),
with again a broad peak in the power spectrum due to the short
observing run. It is interesting to note that the UV light curves
{\em seem} to show an anti-correlated behaviour regarding the
QPOs. However, the UV data sets each cover only one QPO cycle and
therefore cannot serve for a rigorous period analysis. It might be
that the anti-correlated behaviour only appears at the same time as
flares appear.

Fig.~\ref{Figqpohardsoft} shows the soft and hard
X-ray light curves folded on the 5000~s (QPO) period. The variation on
this period is clearly much stronger in the soft component, where we
measure an amplitude of 58\% of the maximum, than in the hard
component, where the amplitude is only about half (30\%) as large.

%-----------------------------------------------------------
   \begin{figure}
   \centering
  \resizebox{8cm}{!}{\rotatebox{270}{\includegraphics{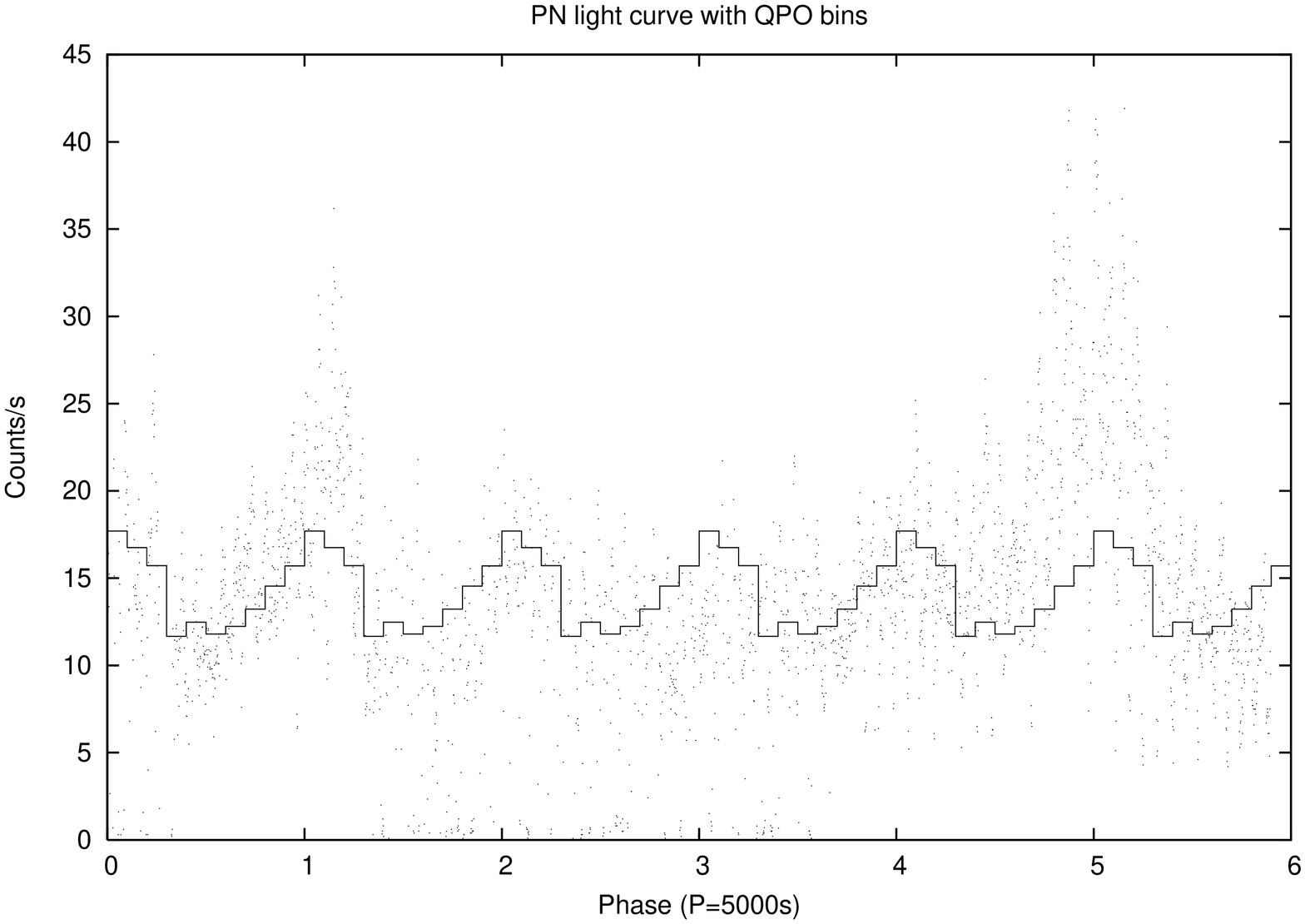}}}

      \caption{PN light curve. Over-plotted is the light curve phased
              on the 5000~s (QPO) period. It is obvious that the
              flares coincide with the maxima in the 5000~s cycle and thus
              dominate the variation.
              }
         \label{Figlcvpnqpo}
   \end{figure}
%
%______________________________________________________________

%%% FLARES CORRELATE WITH MAXIMA IN QPO

Reexamining the flares shows a correlation with a maximum
in the 5000~s (QPO) variation (Fig.~\ref{Figlcvpnqpo}), and thus dominate the
signal. Removing the flares from the light curves reduces the 5000~s (QPO)
signal in the power spectrum significantly by a factor of 4
(comparable in strength to the minor peak at 2000\,s) and shifts it to
approximately~5500\,s. However, the power spectrum for periods
$<1000$\,s is unchanged.

\section{Medium resolution spectroscopy}
\label{mrs}
\begin{figure*} \centering
   \includegraphics[width=8cm]{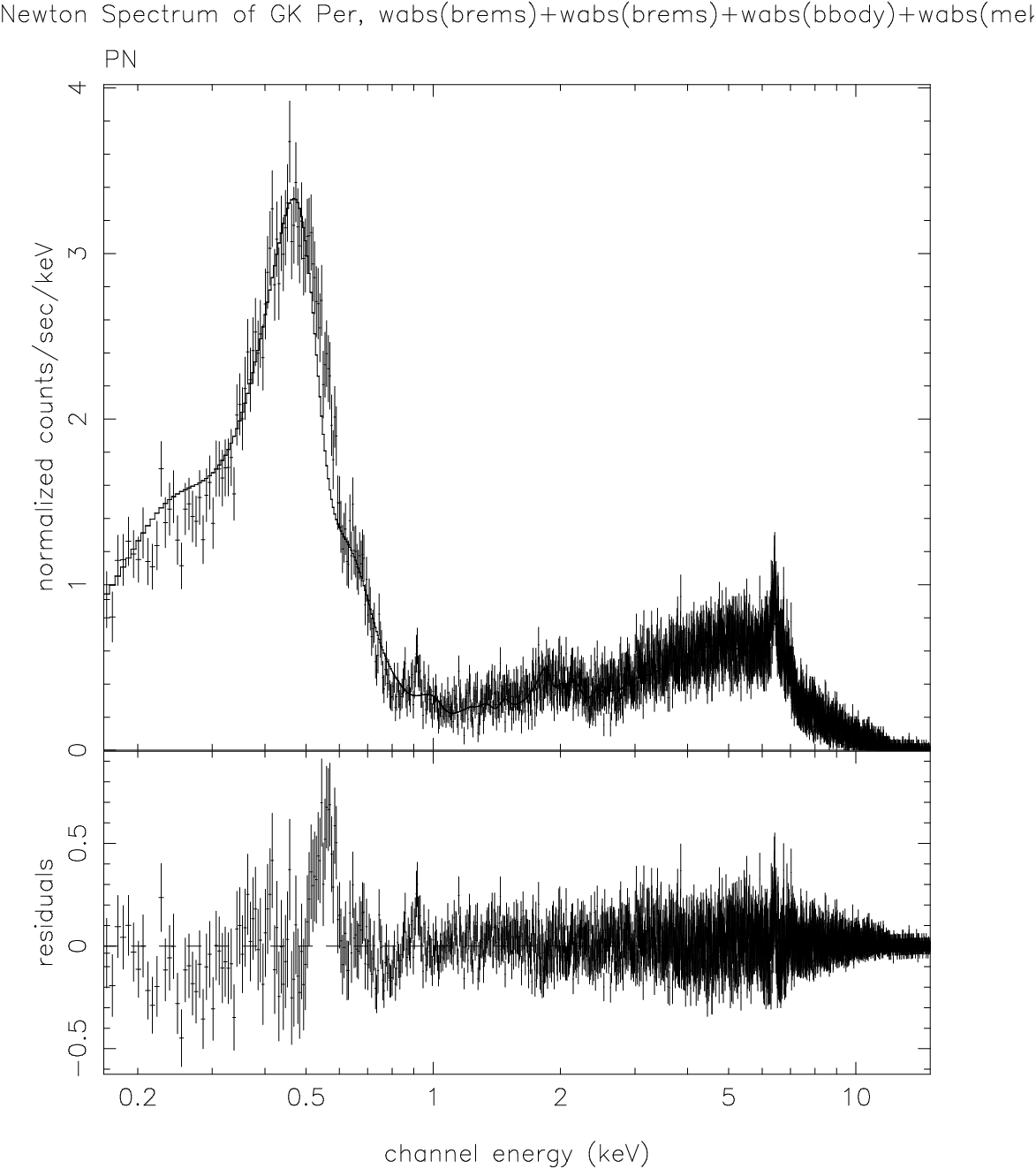}
   \includegraphics[width=8cm]{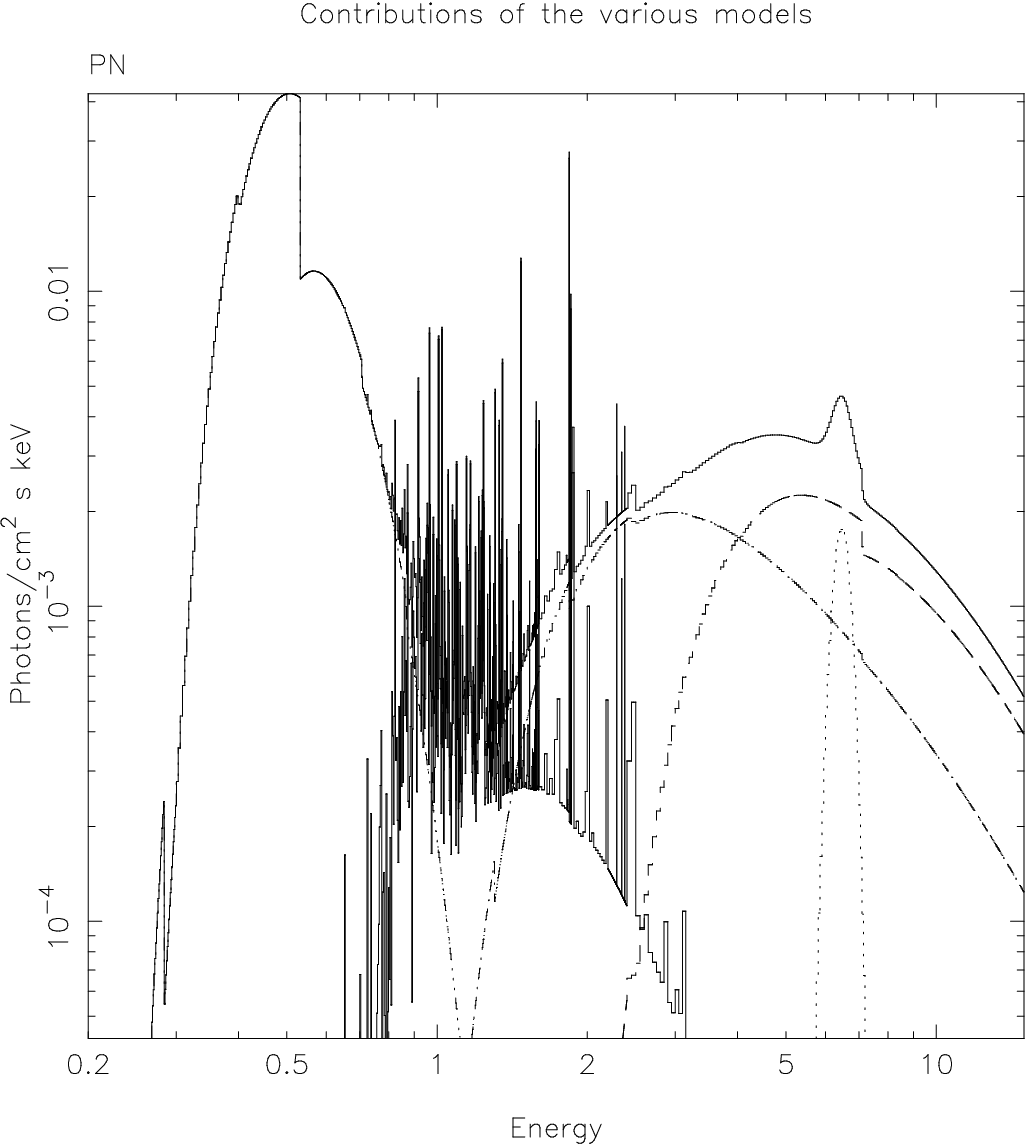} \caption{The
   PN continuum spectrum with a leaky absorber (hard X-rays), a
   blackbody component (soft X-rays), a mekal component and a Gaussian
   (Fe fluorescence 
   at 6.4 keV) spectral fit ($\chi^2 = 0.91$).  } \label{Figpnspec}
   \end{figure*}

Fig.~\ref{Figpnspec} shows the EPIC-PN spectrum of GK Per together
with a fit to the data. Following the approach by Ishida et al. (1992)
we chose a leaky absorber model for the hard X-ray spectrum, and added
a blackbody spectrum for the soft X-ray emission representing the
thermalized emission from the white dwarf and a mekal spectrum for the optically thin emission. The leaky absorber consists
of two bremsstrahlung components with identical temperatures but
different column densities to allow for structured absorption, thus
representing a simple model of a possibly much more complex situation
with a range of column densities due to inhomogeneous in-falling
material. However, this two-component approach represents a simple
blobby accretion model.  On top of
these continuum spectra we added a simple Gaussian to account for the
Fe fluorescence line at 6.4 keV. The resulting fit has a reduced
$\chi^2$ of 0.91.

%%% RESIDUALS
Our model fits the data well, but not perfectly. The fit residuals
(also shown in Fig.~\ref{Figpnspec}) are most prominent in the
region 0.5 to 0.6~keV. This happens to be the spectral region covered by
the RGS instruments and, as shown in Section~\ref{emlines}, we see line
emission (in particular H-like N\,{\sc vii} at $\sim$ 0.9 keV) that is
not appropriately described by the models used. Also, the single
temperature black body is only a simple approximation of a
possibly far more complex temperature structure.

%%% FLUX
The flux over the total observed range in PN (0.17 to 15 keV,
corrected for the missing central pixel) is
$F_{total} = 2.82\times 10^{-10}$\,erg\,cm$^{-2}$\,s$^{-1}$; for
comparison with Hellier et al. (2004) we also compute the flux in the
energy range 2 to 15 keV as $F_{2-15 keV}$ = $2.73 \times 10^{-10}$
erg\,cm$^{-2}$\,s$^{-1}$, i.e., the soft component contributes very
little to the total X-ray flux. Due to the pile-up correction an error
of approximately 10\% is introduced in the flux estimates (Robrade
2005, private communication).
%$F_{1-10 keV} = 1.94 \times 10^{-10}$\,erg\,cm$^{-2}$\,s$^{-1}$.
% comparison of flux
Hellier et al. observed a mean flux during rise and maximum of the 1996
outburst of $F_{2-15 keV; H} = 5.0 \pm 0.8 \times 10^{-10}$\,erg
cm$^{-2}$ s$^{-1}$, similar to Ishida et al.'s (1992) value for the
1989 outburst of $F_{2-20 keV; I} = 4.8 \pm 1.7 \times 10^{-10}$\,erg
cm$^{-2}$ s$^{-1}$.
%luminosity
Using a distance of 460pc our flux value leads to a luminosity of $L_{2-15
keV} = 6.91 \times 10^{33}$\,erg s$^{-1}$ ($L_{total} = 7.14\times
10^{33}$\,erg s$^{-1}$).

Our column densities of the leaky absorber of $(2.20\pm 0.14)
\times10^{23}$\,cm$^{-2}$ and $(3.67\pm0.23)\times10^{22}$\,cm$^{-2}$ are
almost an order of magnitude smaller than those determined by Ishida et
al. (1992) who caught GK Per on early decline from outburst
(approximately 0.5 mag below maximum light).

The leaky absorber model is certainly not the only possible model that
one could fit to the data. Instead of the leaky absorber model we
also tried a warm absorber (single bremsstrahlung component) without
success (unacceptable fit with a $\chi^2 = 12$), a single black body
($\chi^2 = 5.4$) and a single black body plus an optically thin
component (raymond). The latter fit is better but not acceptable ($\chi^2 =
1.75$). Furthermore, physically we
expect bremsstrahlung to originate in the post shock portion of the
accretion column as the material cools while it settles onto the white
dwarf (see e.g.\ King 1995).

The addition of the mekal model improved the fit slightly with a
$\chi^2$ of 0.906 in comparison to the simpler model with a $\chi^2$
of 0.973. An exchange of the mekal model with a raymond model yields
no difference ($\chi^2 = 0.914$) and the parameter values are
identical within the error ranges. In particular, the parameters of
the raymond model lie within the error range of the mekal parameters
(raymond: $n_H = 1.317\pm0.072$; $kT=0.807\pm0.029$; $norm =
0.0107\pm0.0015$).

On the other hand, a more complicated model allowing for a range of
temperatures in the black body component does not seem necessary
regarding the good fits to the data. However, a realistic accretion
column will certainly show a range of temperatures between the shock
and the white dwarf's surface.

%-----------------------------------------------------------
%-----------------------------------------------------------

%__________________________________________________ table
   \begin{table*}
%   \centering
      \caption[]{Fit parameters of the continuum fit (model:
      wabs(blackbody) + wabs(bremsstrahlung) + wabs(bremsstrahlung) + wabs(mekal) +
      gauss), $\chi^2 = 0.906$.

         }
         \label{Tabcontfit2}
%     $$ 
         \hspace{4cm}
         \begin{tabular}{llccc}
            \noalign{\smallskip}
            \hline
            \noalign{\smallskip}
            \hspace*{2ex}Parameter \hspace*{2ex}& \hspace*{2ex}Type of model \hspace*{2ex}& \hspace*{2ex}Unit\hspace*{2ex} & \hspace*{2ex}value \hspace*{2ex}& \hspace*{2ex}error\hspace*{2ex} \\
            \noalign{\smallskip}
            \hline\hline
            \noalign{\smallskip}
            $n_H$ &                & $10^{22}$cm$^{-2}$ & 21.98  & 1.37 \\
            kT    & bremsstrahlung & keV     &  10.56     & 0.81  \\
            norm$^1$  & bremsstrahlung &     & 0.0801  & 0.0052 \\
            \hline
            $n_H$ &                & $10^{22}$cm$^{-2}$ & 3.67    & 0.23 \\
            kT    & bremsstrahlung & keV     &  10.56$^*$      & 0.81$^*$ \\
            norm$^1$  & bremsstrahlung &     & 0.0240  & 0.0018 \\
            \hline
            $n_H$ &                & $10^{22}$cm$^{-2}$ & 0.315  & 0.019 \\
            kT    & black body     & keV     & 0.05961  & 0.00027 \\
            norm$^2$  & black body &         & 0.0119   & 0.0048 \\
            \hline
            $n_H$ &                & $10^{22}$cm$^{-2}$ & 1.306    & 0.068  \\       
            kT    & mekal          & keV     & 0.761    & 0.038  \\
            norm$^3$  & mekal      &         & 0.0112   & 0.0017\\
            \hline
            $E$   & Gaussian       & keV     & 6.475    & 0.015  \\       
         $\sigma$ & Gaussian       & keV     & 0.274    & 0.020  \\
            norm$^4$  & Gaussian   &         & 0.00121  & 0.00014\\

            \hline\hline
         \end{tabular}
%     $$ 

$^*$ value set equal to value of other bremsstrahlung component.

$^1$ norm(brems) = $3.02\times10^{-15} / (4\pi D^2) \int n_e n_I dV$,
where $D$ is the distance to the source (in cm) and $n_e$, $n_I$ are the
electron and ion densities (in cm$^{-3}$).

$^2$ norm(bbody) = $L_{39}/ D_{10}^2$, where $L_{39}$ is the source
luminosity in units of $10^{39}$\,erg s$^{-1}$ and $D_{10}$ is the
distance to the source in units of 10~kpc. With a distance of 460\,pc
this yields a luminosity for the black body component $L = 1.7 \times
10^{33}$~erg s$^{-1}$.

$^3$ norm(mekal) = $(10^{-14}/(4\pi (D_A(1+z))^2)) \int n_e n_H dV $ ,
where $D_A$ is the angular size distance to the source (cm), $n_e$ and
$n_H$ are the electron and H densities (cm$^{-3}$)

$^4$ norm(gaussian) = total photons cm$^{-2}$ s$^{-1}$ in the line.

   \end{table*}
%__________________________________________________ 

%----------------------------------------------------------- 
   \begin{figure} \centering
   \includegraphics[width=7cm]{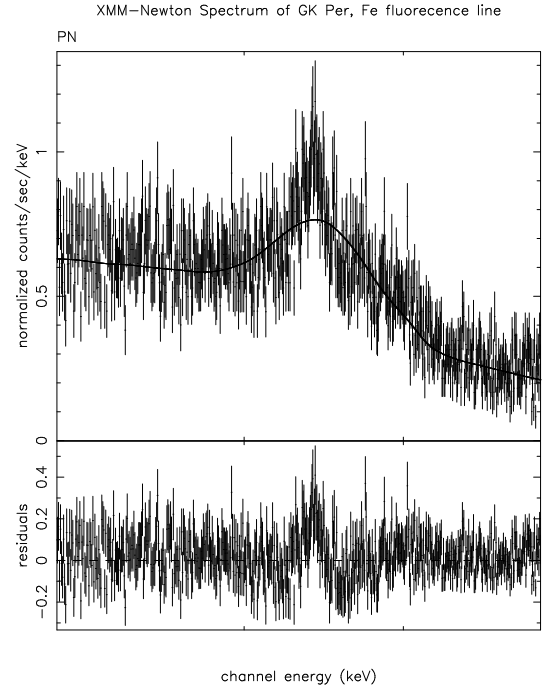}
   \caption{Fe fluorescence line with the model fit (upper panel) and the
     residuals (bottom panel).}  \label{Figfluorescence}
   \end{figure}
%
%______________________________________________________________

The line flux in the Fe fluorescence line is measured as $1.25 \times
10^{-11}$\,erg\,cm$^{-2}$\,s$^{-1}$ ($1.21\times 10^{-3}$
photons\,cm$^{-2}$\,s$^{-1}$), a factor two less than determined by
Ishida et al. (1992) during the 1989 moderate outburst. The line has
an equivalent width of 447 eV, considerably more than in Hellier \&
Mukai's (2004) observations who measure an equivalent width of 260
eV. Closer inspection of Fig.~\ref{Figfluorescence} suggests that our
value might be overestimated as the single Gaussian fit to the Fe
fluorescent line is not satisfactory: the center of the line seems
narrower than the fitted Gaussian and there seems to exist a broad
underlying component, especially redwards of the line. This emission
in the red wing of the fluorescence line might be Compton scattering
due to K-shell fluorescence in the white dwarf atmosphere as discussed
by van Teeseling et al. (1996). In this case, fluorescence in the
accretion column would be less significant.  However, Hellier \& Mukai
(2004) attribute this emission to infalling pre-shock material with
velocities up to 3700\,km\,s$^{-1}$.

% no variation of Fe fluorescence line on spin and QPO
Interestingly, the variation of the Fe fluorescence line on the spin
cycle is relatively weak. While the total PN light curve shows a
variation of 39\%, the Fe fluorescence line light curve (including
continuum) shows a variation of only 28\% (Fig.~\ref{Figspinlcv}).
Considering the strong continuum at this wavelength (about 2/3 of the
total flux at this wavelength) this amounts to a
negligible variation of the fluorescence line. On the 5000~s (QPO) period the
variation is also negligible.

% spectra for spin high and low phase

In order to verify a change in the amount of absorbing material as a function of
spin cycle we extracted PN-data for high and low spin phases and fitted
both spectra with the leaky absorber plus black body model, keeping
all parameters fixed except the column densities.
The resulting fits with $\chi^2$ values of 0.86 and 0.81 are
better than the original fit to the spin cycle averaged data
($\chi^2 = 0.91$). The column densities change significantly and are
much better defined in the separate fits -- as implied by the smaller errors
(see Table~\ref{TabnH}).

%__________________________________________________ table
   \begin{table*}
   \centering
      \caption[]{Fitted column densities (in units $10^{22}$cm$^{-2}$) 
      and equivalent width (EW) of the fluorescence line (in eV) for the
      averaged spectrum (see Table~\ref{Tabcontfit2}) as well as
      for the spin and 5000~s (QPO) cycle high and low states. The
      given flux is in units of $10^{10}$\,erg\,cm$^{-2}$\,s$^{-1}$.
         }
         \label{TabnH}
%     $$ 
         \begin{tabular}{lcccccccccc}
            \noalign{\smallskip}
            \hline
            \noalign{\smallskip}
            model & average & $\sigma$ & high state & $\sigma$ &
            low state & $\sigma$ & QPO high state & $\sigma$ & QPO low state & $\sigma$ \\
            \noalign{\smallskip}
            \hline\hline
            \noalign{\smallskip}
            brems & 21.98  & 1.37   & 17.54  & 0.29   & 32.54  & 0.61   & 19.69  & 0.33   & 26.31   & 0.48   \\
            brems & 3.67   & 0.23   & 3.049  & 0.055  & 5.180  & 0.104  & 2.720  & 0.047  & 5.93    & 0.13   \\
            bbody & 0.315  & 0.019  & 0.2954 & 0.0013 & 0.3435 & 0.0018 & 0.2809 & 0.0012 & 0.3746  & 0.0021 \\
            mekal & 1.306  & 0.068  & 1.295  & 0.026  & 1.432  & 0.029  & 1.231  & 0.023  & 1.581   & 0.037  \\ \hline
            $\chi^2$ & 0.91 & & 0.86 & & 0.81 & & 0.80 & & 0.80 & \\
            Flux     & 2.82 & & 3.03 & & 2.45 & & 2.98 & & 2.57 & \\
            EW       &  447 & &  417 & &  523 & &  430 & &  483 & \\
            \hline\hline
         \end{tabular}
%     $$ 
   \end{table*}
%__________________________________________________ 

% spectra for QPO high and low phase

Similarly, we extracted the spectra during high (flare) and low
phase of the 5000~s (QPO) variation.
These changes between (QPO) high and low state can also solely be
explained by a variation in the column density. The column densities
are listed in Table~\ref{TabnH}.  It is interesting to note that the
variation in the column density during the spin cycle is strongest in
the bremsstrahlung component and during the 5000~s (QPO) cycle (the flares)
strongest in the black body component.

\section{High-resolution spectroscopy}
\label{emlines}

%-----------------------------------------------------------
   \begin{figure*} \centering
   \includegraphics[width=8cm]{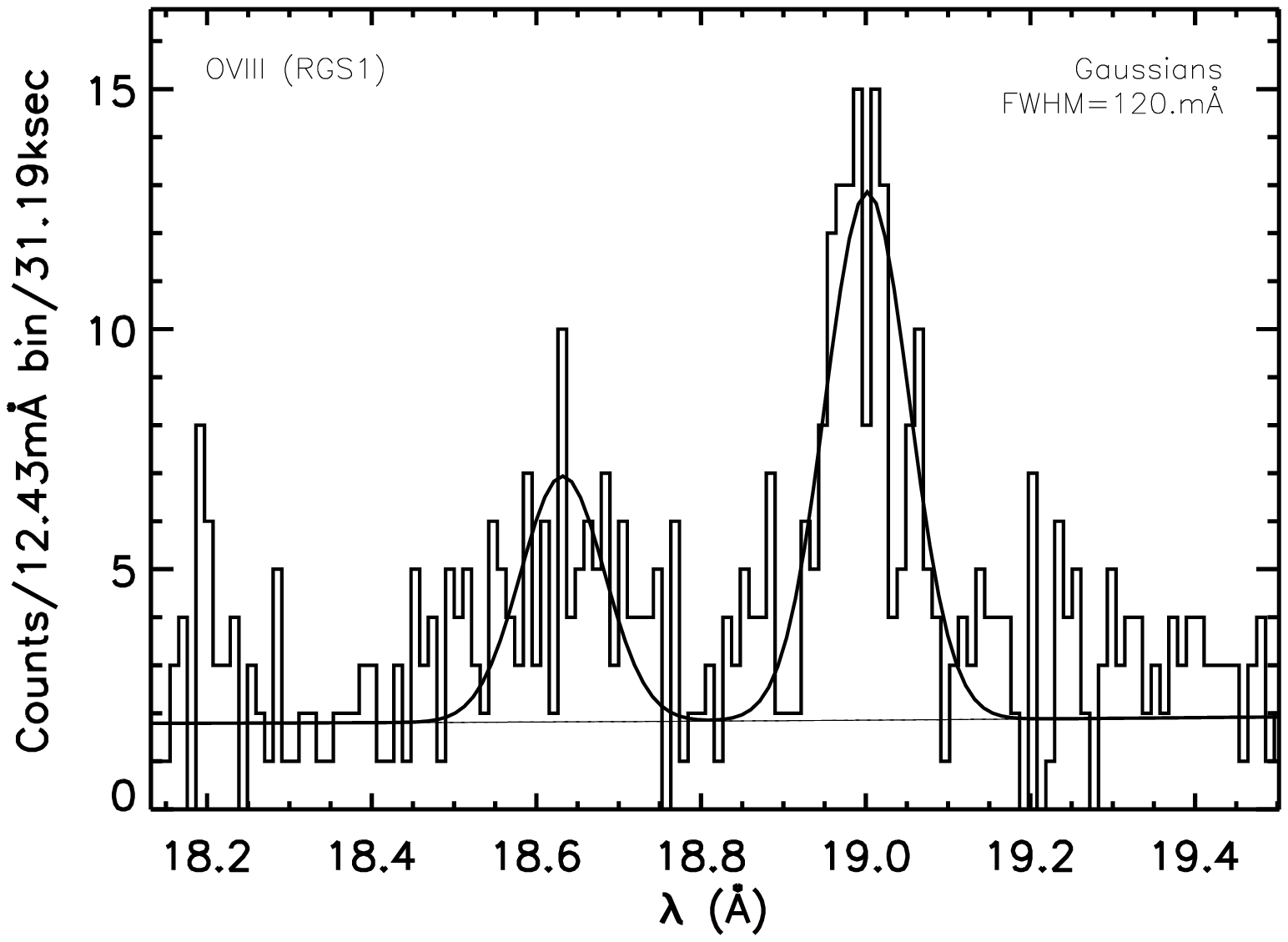}
   \includegraphics[width=8cm]{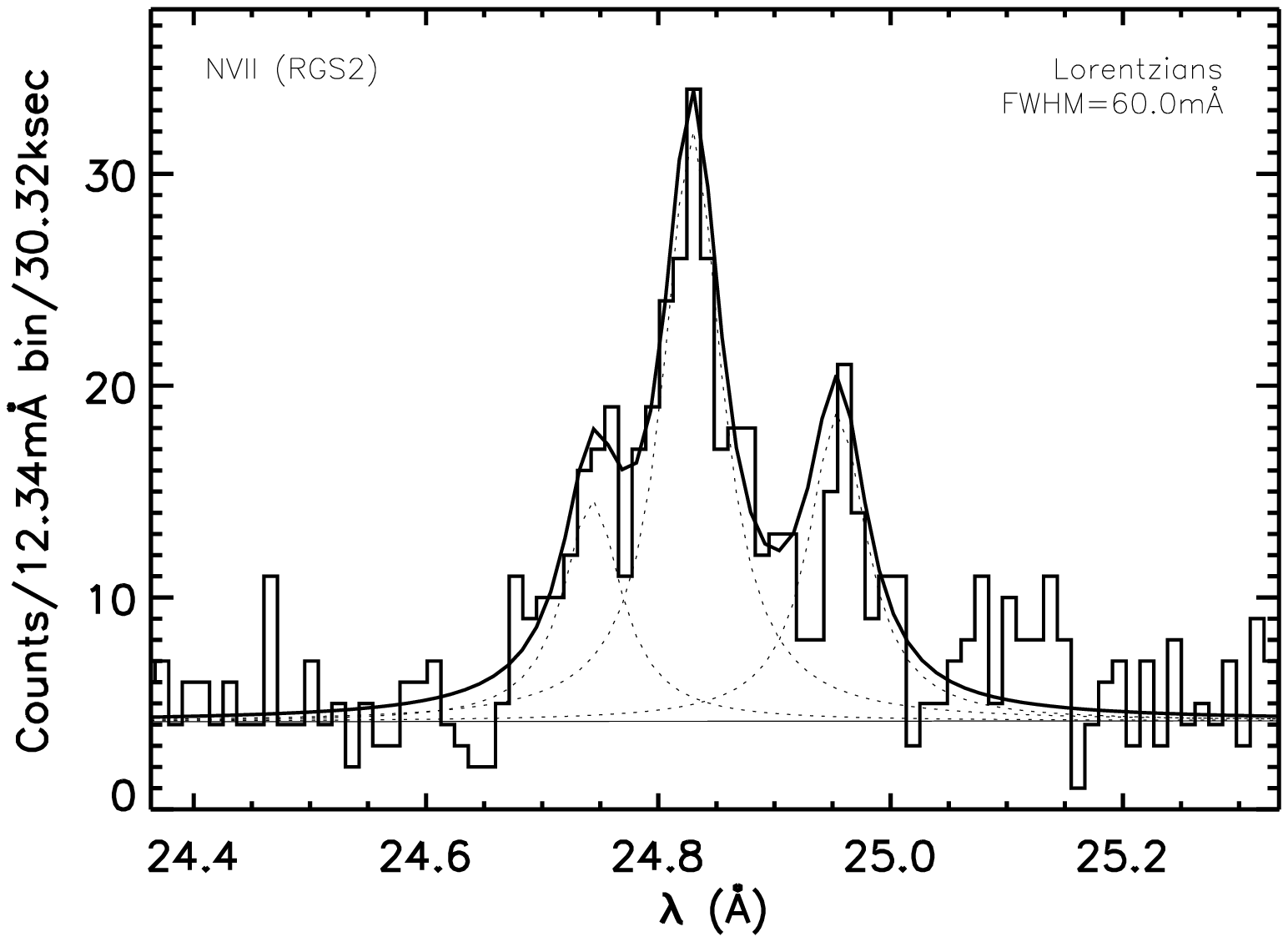}
   \includegraphics[width=8cm]{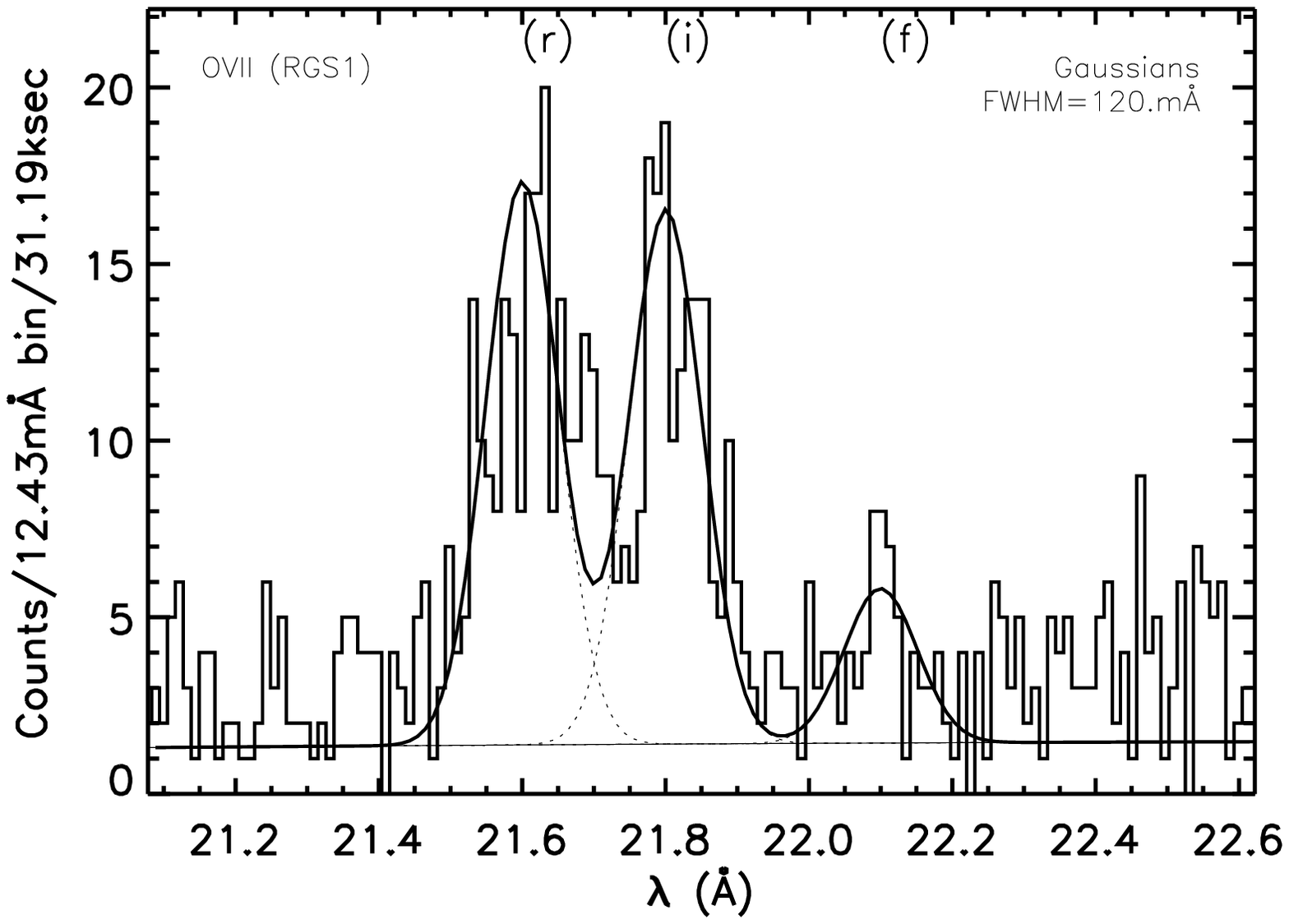}
   \includegraphics[width=8cm]{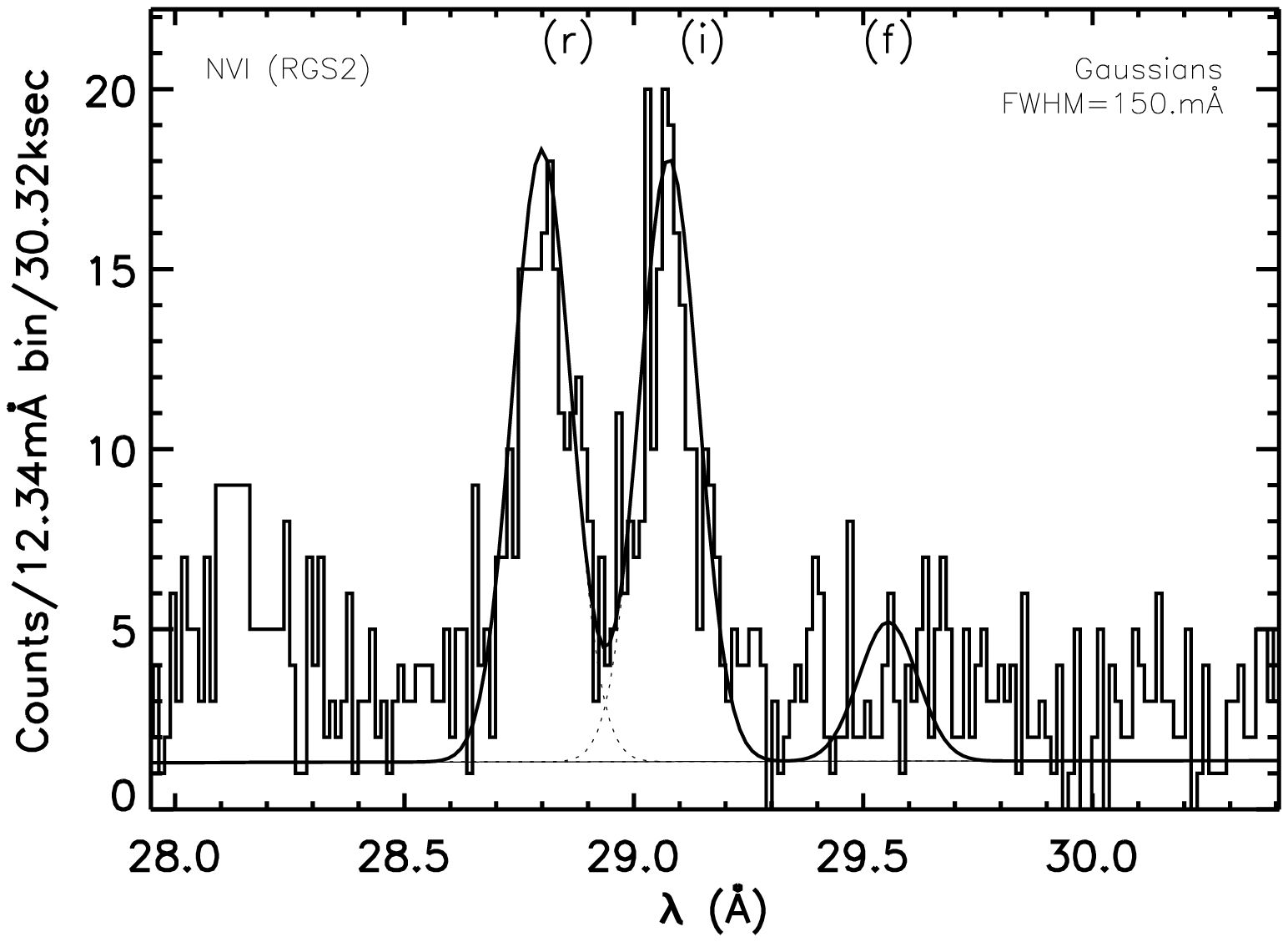}

      \caption{O\,{\sc viii}, N\,{\sc vii} (H-like) and O\,{\sc vii}, N\,{\sc vi}
              (He-like) emission lines in RGS1 (left) and RGS2 (right). The FWHM
	      values are denoted by $\sigma$ in the figures.
              }
         \label{Figonlines}
   \end{figure*}
%
%______________________________________________________________
%

%-----------------------------------------------------------
   \begin{figure}
   \centering
  \includegraphics[width=8cm]{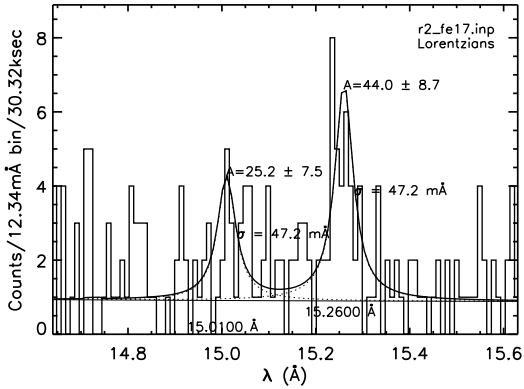}

      \caption{Fe\,{\sc xvii} emission line. Note, that the 15.26\, \AA\ 
              intersystem line is stronger than the 15.01\, \AA\ 
              resonance line.
              }
         \label{Figfe17}
   \end{figure}
%
%______________________________________________________________
%
\subsection{H-like and He-like emission lines}
\label{emls}
The {\it XMM-Newton} reflection grating spectrometers cover the
energy range between 0.3 - 2 keV, where the overall X-ray flux
in the medium resolution spectrum is lowest.  Despite the overall
rather low signal-to-noise ratio of the RGS data, a multitude of
emission lines can be detected. The line features
can be identified as O\,{\sc viii}, N\,{\sc vii} (both H-like) as well
as O\,{\sc vii} and N\,{\sc vi} (both He-like); in order to give
an impression of the RGS data we plot the most
prominent emission lines in the RGS1 and RGS2 spectra in 
Fig.\ref{Figonlines}.

In Table~\ref{Tablines} we list the line counts, line widths and
identifications of other He-like and H-like lines found in the
spectra.  The line fluxes were obtained with the CORA program (Ness \&
Wichmann 2002). The line profiles were approximated by analytical
profile functions, where the Full Width at Half Maximum (FWHM) can be
iterated. We found that all lines are significantly broader than the
instrumental broadening profiles (FWHM$=0.06$\,\AA), so physical line
broadening mechanisms are required. The instrumental broadening can
roughly be described by a Lorentzian profile. However, for the
broadened lines of O\,{\sc viii}, O\,{\sc vii} and N\,{\sc vi} we
found Gaussian profiles to fit the RGS data much better than
Lorentzians. The FWHM values increase with increasing wavelength
indicating similar velocities for these lines ($\sim
1600-1800$\,km\,s$^{-1}$).  The N\,{\sc vii} line at 24.8\,\AA\ shows
a totally different profile with three obvious components (upper
right in Fig.\ref{Figonlines}. We approximated these by Lorentzian
profiles (since they are more affected by the instrumental line
broadening).\\

%The RGS has the great advantage of providing two independent grating
%spectra, however, it is not always possible to fully benefit from
%this, because the CCDs on which the dispersed photons are recorded,
%suffered severe failures during the past years. The N\,{\sc vii} and
%N\,{\sc vi} lines can only be measured with the RGS2 and the O\,{\sc
%vii} only with the RGS1. So, the triple nature of the N\,{\sc vii}
%line can neither be confirmed nor rejected by the comparison. Only the
%O\,{\sc viii} line can be measured with both instruments and the
%results are consistent.\\

\subsection{Fe\,{\sc xvii} emission lines}

Let us focus next on the Fe\,{\sc xvii} lines at 15.01\,\AA\ and
15.26\,\AA, shown in Fig.\ref{Figfe17}. These lines are fitted best with
Lorentzian profiles.  The line at 15.26\,\AA\ is clearly stronger than
the 15.01-\AA\ line, an interesting finding since the 15.01-\AA\ line
is the strongest resonance line of the Fe\,{\sc xvii} species
(2p$^6$\,$^1$S$_0$--2p$^5$\,3d\,$^1$P$_1$) involving high transition
probabilities, while the 15.26-\AA\ line is an intersystem line
(2p$^6\,^1$S$_0$--2p$^5$3d\,$^3$D$_1$) with correspondingly lower
transition probabilities. Of course, the identification might be
incorrect and we are actually dealing with the O\,{\sc viii}
Ly$_\gamma$-line expected at 15.17\AA.  However,
%two considerations
%make this possibility unlikely. First, 
the interpretation as O\,{\sc viii} Ly$_\gamma$ implies a red shift,
which we do not see in
the other lines of the O\,{\sc viii} Ly series nor in any other
line.
% Second, the line flux (44 counts) is significantly higher than
%that found in the Ly$_\beta$-line at 16\,\AA\ (24 counts). While this
%could be explained by scattering processes affecting the low-n Ly
%lines stronger than the higher-n Ly lines, this scenario is ruled out
%by the strength of the Ly$_\alpha$ line at 19\,\AA\ (upper left in
%Fig.\ref{Figonlines}). 
We therefore conclude that the line measured
with 44 counts must be attributed to the 15.26-\AA\ line of Fe\,{\sc
xvii}, while the 15.01-\AA\ line is only measured with at most 25 counts. The
errors are quite large, but the ratio of $\lambda 15.01/\lambda15.26$
is, within the errors, less than unity.

Since both lines originate from the same ionization stage of the same
element, this cannot be attributed to temperature or
abundance effects. At any temperature capable of producing the ionization
stage of Fe\,{\sc xvii} the ratio is expected to be of order 3-4. The
only scenario explaining the measured line ratio is resonant
scattering, i.e., reabsorption of photons emitted from transitions with
high radiative absorption probabilities and reemission into any other
direction. In symmetrical emission regions the effect balances out by
some photons being scattered out of the line of sight and others being
scattered into the line of sight. However, in asymmetric geometries
like accretion discs and curtains, resonant line scattering can lead
to an effective loss of resonance line photons. Lines with low
oscillator strengths (like the 15.26-\AA\ line) are much less affected
by resonance scattering.

We tested this scenario by searching for other lines of Fe\,{\sc xvii}
with low oscillator strength, and found another line at 16.78\,\AA\
(2p$^6\,^1$S$_0$--2p$^5$3s$^3$P$_1$) to be suspiciously strong
compared to the 15.01-\AA\ line. This supports the above explanation
and we therefore conclude that the regions emitting Fe\,{\sc xvii}
come from asymmetric plasma with high densities or long path lengths.
We further note that the oxygen emission cannot originate from the
same regions, because the same effect would have to be observed in an
absence of the Ly$_\alpha$-line and a stronger presence of the
Ly$_\beta$-line, which is not the case. One could argue that the
Ly$_\alpha$ {\em and} the Ly$_\beta$-lines are affected by resonant
scattering while the Ly$_\gamma$-line is not (and it would be the true
interpretation of the 15.26\,\AA\ feature). However, this also
requires resonant scattering to be at work (only for oxygen instead of
iron), but would still not explain the red shift and the relative
strength of the 16.78-\AA\ line. We thus consider it most likely that
iron is emitted in a different geometry (where resonance photons are
lost) than oxygen (regions which are either optically thin or are more
symmetric).\\

\subsection{Densities}

He-like triplets can be used for plasma density diagnostics using the
density-sensitive f/i line flux ratios.  This technique was
originally developed by Gabriel \& Jordan (1969) for solar X-ray data
and has recently been applied by, e.g., Ness et al. (2004) to analyze
a large number of stellar coronal emission line spectra.  In the bottom left
of Fig.~\ref{Figonlines} we show the O\,{\sc vii} triplet with the
triplet lines marked by r (resonance line), i (intercombination line)
and f (forbidden line).  The forbidden line is quite faint, while the
intercombination line has almost the same flux as the resonance line
(within the errors).  In plasma with an electron density below
$10^{11}$\,cm$^{-3}$ the line flux ratio f/i is greater than one. From
the measured ratio f/i$\sim 0.3$ we formally derive a density of $\log
n_e\sim 11.6\,\pm\,0.1$\,cm$^{-3}$. This number assumes negligible
radiative deexcitations of the forbidden line level.  However, in the
presence of UV radiation fields, the transition from f to i can also
be triggered by photons and thus mimic high densities. We have performed
a crude estimate of these effects using the UV flux measurements by Wu
et al. (1989) and Yi \& Kenyon (1997), who find a very flat UV
spectrum with $\log$ Flux (erg\,cm$^{-2}$\,s$^{-1}$\,\AA$^{-1}$) $\approx$
-11.75 over the range 1000 to 3500\,\AA\ .  As in Ness et
al. (2001) we convert this flux to a black-body temperature, which can
be used to predict a correction term in the density determination. We
find a corresponding black-body temperature of $< 2000$\,K assuming a
wavelength for the transition f to i of 1636\,\AA. With such a low
temperature any effects from the UV radiation field can be neglected
for all He-like ions. We additionally use the N\,{\sc vi} and Ne\,{\sc
ix} He-like triplets and find $\log n_e = 11.57\pm0.1$ for N\,{\sc vi}
and $\log n_e = 12.1\,\pm\,0.3$ for Ne\,{\sc ix}.

The densities and the line flux ratio $\lambda 15.01/\lambda 15.26$
can be explored with a few assumptions to estimate the path length
$\ell$. First we assume a simple "escape factor" model (escape
factor $P(\tau)\approx [1+0.43\tau]^{-1}$ for a homogeneous mixture of
emitters and absorber in a slab geometry; e.g., Kaastra
\& Mewe 1995, Mewe et al. 2001). The escape factor can be calculated
from the measured line flux ratio and the line flux ratio in an
optically thin plasma (i.e., the unabsorbed line flux ratio). The
unabsorbed ratio is somewhat uncertain with values up to
four (from theoretical calculations; e.g., Brown et al. 1998, Bhatia
\& Doschek 1993) and values in the range 2.8-3.2 (from laboratory
measurements in the Livermore Electron Beam Ion Trap EBIT). Systematic
measurements of the $\lambda 15.01/\lambda 15.26$ ratios have been
carried for a large sample of stellar coronae by Ness et al. (2003)
yielding an average value of $\sim 3$, which we use here. The formal
measured ratio is $\lambda 15.01/\lambda 15.26\approx 0.6$ and the
escape factor $P(\tau)\approx 0.2$.  The escape factor model allows us
to calculate an optical depth $\tau\approx 10$.

The equation
\begin{equation}
\label{tau}
\tau=1.2\
10^{-14}\left(\frac{n_i}{n_{el}}\right)A_z\left(\frac{n_H}{n_e}\right)\lambda\
%f\sqrt{\frac{M}{T}} {\red n_e}\ell
f\sqrt{\frac{M}{T}} {n_e}\ell
\end{equation}
(e.g., Mewe et al. 2001) describes the optical depth $\tau$ as a function of some
ion-specific
parameters and the product of electron density $n_e$ (cm$^{-3}$) and
path length $\ell$ (cm).  Further, $n_i/n_{el}$ is the fractional
ionisation, $A_z=n_{\rm el}/n_{\rm H}$ is the elemental abundance,
$n_H/n_e=0.85$ is the ratio of hydrogen to electron density, $f$ is
the oscillator strength, $M$ is the atomic number, $\lambda$ is the
wavelength of the resonance line (\AA) and $T$ is the temperature
(K). From the optical depth $\tau\approx 10$, derived from $\lambda
15.01/\lambda 15.26$ line ratio, a characteristic plasma dimension
$\ell$ can be estimated for the $\lambda=15.01$-\AA\ line. Using
$M=26$ (for iron), $f=2.6$ (for the $^1$S$_0$--$^1$P$_1$ transition at
15.01\,\AA), $T=7\times 10^6$\,K.  We further assume an ionisation
fraction $n_i/n_{el}=0.25$ in Fe\,{\sc xvii} (Arnaud \& Raymond 1992)
%fraction $n_{FeXVII}/n_{Fe}=0.25$ in Fe\,{\sc xvii} (Arnaud \& Raymond 1992)
and cosmic Fe abundance. With these parameters we compute from
Eq.~\ref{tau} and the measured value $\tau\approx 10$ the product
$n_e\ell=1.3\times 10^{21}$\,cm$^{-2}$.  The electron density $n_e$
can only be roughly estimated from the He-like triplet diagnostics,
which consistently yields values $\log n_e \sim 11.8$. Note, however,
that the iron lines might originate from totally different regions
than the He-like lines and these density estimates might not be
valid. With this value we find a formal path length of $\ell =2\times
10^9$\,cm with considerable uncertainty.

\subsection{Variation of emission lines?}

An extraction of line profiles of O\,{\sc viii}, N\,{\sc vii} and
N\,{\sc vi} for high and low, pole approaching or pole receding spin
phases shows no significant variation (no detectable line strength
variation or line shifts) with spin cycle. These emission lines also
show no significant variation on the 5000~s (QPO) cycle (or rather with the
flares) with the given quality of the RGS spectra.

%___________________________________________________________________ table

   \begin{table} \caption[]{Lines identified in RGS spectra.}
      \label{Tablines}
\vspace{-.6cm}

	$$
         \begin{array}{llcrclrcl}
            \noalign{\smallskip}
            \hline
            \noalign{\smallskip}
            Line{\hspace{4ex}} & {\hspace*{2ex}}\lambda_0~(\AA){\hspace{2ex}} & RGS &
            {\hspace{2ex}}Counts & & FWHM &
             {\hspace{2ex}}\lambda~(\AA) &
             {\hspace{1ex}}T_{max} \\
            \noalign{\smallskip}
            \hline
            \noalign{\smallskip}

\multicolumn{4}{l}{\rm \bf H-like\ ions}&&(\AA)&\\\hline

{\mathrm{Si XIV}} & 6.180   & 2 & 48\,\pm\,11   & & 0.047^a  &  6.1800 &  7.11    \\
\hline
{\mathrm{Ne X}} & 12.132     & 2 & 73\,\pm\,12   & & 0.08^a   & 12.16 & 6.76   \\
\hline
{\mathrm{O VIII}} & 16.003   & 1 &  25\,\pm\,\ 7    & & 0.05^a  & 16.003  & 6.50   \\
{\mathrm{O VIII}} &          & 2 &  30\,\pm\,\ 8    & & 0.05^a & 16.003  &        \\
\hline

{\mathrm{O VIII}} & 18.971/ & 1 & 127\,\pm\,13   & & 0.12^b   & 19.003 & 6.48/9 \\
                  & \ \ 18.977  & 2 & 186\,\pm\,17   & & 0.12^b   & 19.0019 &        &        \\
\hline

{\mathrm{N VII}} & 24.779/ & 2 &  85\,\pm\,16   & & 0.06^a   & 24.7431 &  6.32   \\
{\mathrm{N VII}} & \ \ 24.785    & 2 & 211\,\pm\,21   & & 0.06^a   & 24.8294 &         \\ 
{\mathrm{N VII}} &               & 2 & 128\,\pm\,20   & & 0.06^a   & 24.9544 &         \\ \hline
{\mathrm{N VII}} & total         & 2 & 424\,\pm\,31   & &         & 24.8423 &         \\

\hline\hline

\multicolumn{7}{l}{\rm \bf He-like\ ions}\\\hline

{\mathrm{Si XIII}} & 6.648   & 2 & 43\,\pm\,13   & & 0.047^a  &  6.6500 &  7.01   \\
{\mathrm{Si XIII}} & 6.688   & 2 &  2\,\pm\,\ 2   & & 0.047^a  &  6.6900 &  6.99   \\ 
{\mathrm{Si XIII}} & 6.740   & 2 & 21\,\pm\,10   & & 0.047^a  &  6.7400 &  7.00   \\ \hline
{\mathrm{Si XIII}} & total   & 2 & 66\,\pm\,19   & &         &         &  7.03   \\

\hline

{\mathrm{Mg XI}} &  9.170    & 2 & 15\,\pm\,\ 8   & & 0.047^a  &  9.1700 &  6.81   \\
{\mathrm{Mg XI}} &  9.232    & 2 &  9\,\pm\,\ 7   & & 0.047^a  &  9.2300 &  6.80   \\
{\mathrm{Mg XI}} &  9.315    & 2 & 27\,\pm\,\ 8   & & 0.047^a  &  9.3200 &  6.81   \\ \hline

{\mathrm{Mg XI}} &  total    & 2 & 50\,\pm\,13   & &         &         &  6.81   \\

\hline

{\mathrm{Ne IX}} & 13.448    & 2 & 42\,\pm\,\ 9   & & 0.04^a   & 13.4500 &  6.59   \\
{\mathrm{Ne IX}} & 13.553    & 2 & 51\,\pm\,10   & & 0.04^a   & 13.5500 &  6.58   \\
{\mathrm{Ne IX}} & 13.700    & 2 & 53\,\pm\,10   & & 0.04^a   & 13.7000 &  6.59   \\ \hline

{\mathrm{Ne IX}} & total     & 2 & 146\,\pm\,17  & &         &         &  6.59   \\

\hline
{\mathrm{O VII}} & 18.628    & 1 & 62\,\pm\,10   & & 0.12^b   & 18.6318 &  6.34   \\
{\mathrm{O VII}} & 18.628    & 2 & 71\,\pm\,11   & & 0.12^b   & 18.6318 &          &       \\

\hline

{\mathrm{O VII}} & 21.602    & 1 & 186\,\pm\,15   & & 0.13^b   & 21.600  & 6.33   \\
{\mathrm{O VII}} & 21.800    & 1 & 170\,\pm\,14   & & 0.13^b   & 21.800  & 6.32   \\
{\mathrm{O VII}} & 22.100    & 1 &  50\,\pm\,\ 9   & & 0.13^b   & 22.100  & 6.32   \\\hline

{\mathrm{O VII}} & total     & 1 & 406\,\pm\,22   & &         &         & 6.32   \\

\hline

%{\mathrm{N VI}}  & 23.30     & 1 &  36\,\pm\,11   & & 0.15^b   & 23.33   & 6.19   \\
%                 & 23.770    & 1 &  66\,\pm\,12   & & 0.15^b   & 23.7700 & 6.19    & \\ \hline

{\mathrm{N VI}}  & 28.790    & 1 & 178\,\pm\,16   & & 0.15^b   & 28.8500 &  6.17   \\
{\mathrm{N VI}}  & 29.090    & 1 & 120\,\pm\,13   & & 0.15^b   & 29.1145 &  6.16  \\
{\mathrm{N VI}}  & 29.530    & 1 &  27\,\pm\,\ 8   & & 0.15^b   & 29.5545 &  6.16  \\ \hline

{\mathrm{N VI}}  & total     & 1 & 325\,\pm\,37   & &         &         &  6.16  \\ \hline
{\mathrm{N VI}}  & 28.790    & 2 & 202\,\pm\,16   & & 0.15^b   & 28.7993 &  6.17   \\
{\mathrm{N VI}}  & 29.090    & 2 & 197\,\pm\,15   & & 0.15^b   & 29.0766 &  6.16  \\
{\mathrm{N VI}}  & 29.530    & 2 &  46\,\pm\,\ 9   & & 0.15^b   & 29.5545 &  6.16  \\ \hline

{\mathrm{N VI}}  & total     & 2 & 455\,\pm\,40   & &         &         &  6.16  \\

\hline\hline

{\mathrm{Fe XVII}} & 15.014  & 2 & 25\,\pm\,\ 8   & & 0.047^a  & 15.01   & 6.72   \\
{\mathrm{Fe XVII}} & 15.265  & 2 & 44\,\pm\,\ 9   & & 0.047^a  & 15.26   & 6.72   \\ \hline
{\mathrm{Fe XVII}} & total   & 2 & 69\,\pm\,12   & & 0.047^a  & 15.26   & 6.72  \\

%            \noalign{\smallskip}
            \hline\hline
\multicolumn{7}{l}{^a{\rm Lorentzian\,\, profile}\hspace{1cm}^b{\rm Gaussian\,\, profile}}
         \end{array}
     $$
   \end{table}
%__________________________________________________ 

\section{The quiescent data}

Fig.~\ref{Figshell} shows the nova shell clearly visible at X-ray
wavelengths predominantly at soft X-ray wavelengths (up to 0.9 keV).
In comparison to Balman \& \"Ogelman (1999)'s ROSAT data (0.1-2.4 keV)
the shell in our image appears somewhat more extended and the south
western arc is more homogeneous while still showing structured
emission.  Also the radio images obtained in 2002 and 2003 follow this
pattern (Anupama \& Kantharia 2005). In contrast Anupama \&
Kantharia's optical images taken in 2003 show a nearly radially
symmetric shell structure. The X-ray emission coincides with the
(south western) outer edge of the optical shell and Anupama \&
Kantharia deduce that the shell is in shock interaction with the
ambient medium, in particular in the south western direction. Balman
(2002a,b) and Balman (2005) report detailed studies of the same
Chandra observations of this nova shell, in particular they find
significant enhancement of Neon and Nitrogen in the shell in
comparison to solar values.

%----------------------------------------------------------- 
   \begin{figure}
   \centering
   \includegraphics[width=7cm]{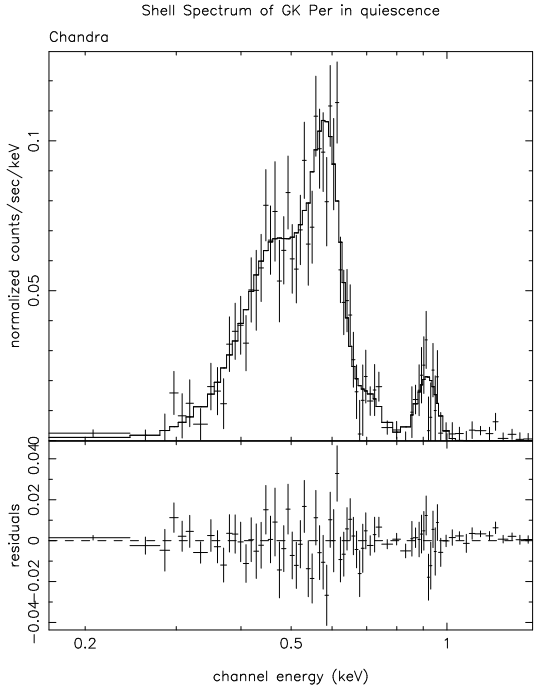}
      \caption{ACIS-S nova shell spectrum of GK Per for the emission shocked
   material (in the south western direction from the central
   object). The fitted model (solid line) includes only pure line
   emission (see Table~\ref{Tabqquiesfit}).}
         \label{Figshellspec}
   \end{figure}
%
%______________________________________________________________

Fig.~\ref{Figshellspec} shows the {\it Chandra} ACIS-S shell spectrum
of GK Per. At energies above 1.5\,keV there is no significant shell
emission.  The shell spectrum can be satisfactorily fit (reduced
$\chi^2 = 1.13$) with a pure emission line spectrum, e.g. with the
He-like emission line series from N\,{\sc vi} (420-532 keV), O\,{\sc
vii} (561-713 keV) and Ne\,{\sc ix} (904-922 keV)
(Tab.~\ref{Tabqquiesfit}) although the photon energies are very
uncertain. For example, Ne\,{\sc ix} could be replaced by Fe\,{\sc
xix} as these lines could not be resolved with ACIS-S. However, the
lower temperature necessary for Ne\,{\sc ix} makes its lines more
plausible.

Similarly to the outburst spectra, the quiescent spectrum of GK Per
can be fitted with a leaky absorber model in addition to an optically
thin raymond model (instead of the black body spectrum). Using two
instead of one bremsstrahlung component significantly improved the fit
(the $\chi^2$ reduced from 1.27 to 1.12), especially in the soft X-ray
region. The formally best fit
($\chi^2 = 1.09$) to the {\it ACIS-S} data is obtained with a
bremsstrahlung temperature of 197 keV, which we consider
unrealistically high; therefore we fixed the bremsstrahlung
temperature to the same value as in the outburst spectra.

The strongest of the shell emission lines (O\,{\sc vii}
573.95 keV) was included for an improved fit of the soft
end of the X-ray spectrum, it might be contamination from the shell
spectrum in in the line of sight. As in the outburst data we had to fix the
bremsstrahlung temperature.  A comparison to Norton et al.'s (1988)
fit to their quiescent spectra shows that they also find a very high,
not well constrained bremsstrahlung temperature of $33^{+95}_{-11}$
keV in one of their data sets.  A comparison of
Table~\ref{Tabqcontfit} with Table~\ref{Tabcontfit2} shows that the
main difference between the two leaky absorber models lies in the
column density of the more strongly absorbed portion of the
bremsstrahlung components and the $norm$ value. As noted below the
table, the $norm$ value depends on the electron and ion densities
$n_e$ and $n_I$ as well as the volume of the emitting material.

%----------------------------------------------------------- 
   \begin{figure}
   \centering
   \includegraphics[width=7cm]{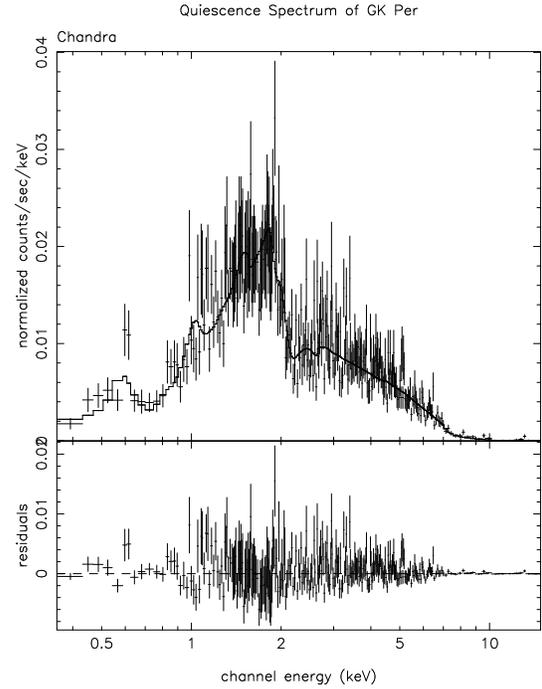}
      \caption{ACIS-S quiescent spectrum of GK Per with a leaky absorber
   plus raymond fit and additional OVII emission.}
         \label{Figquiesspec}
   \end{figure}
%
%______________________________________________________________
 %______________________________________________________________
   \begin{table*}
   \centering
      \caption[]{Fitted emission lines to the quiescent shell spectrum
   (ACIS-S).
         }
         \label{Tabqquiesfit}
%     $$ 
%         \hspace{2cm}
         \begin{tabular}{lcccc}
            \noalign{\smallskip}
            \hline
            \noalign{\smallskip}
	    Ion \hspace*{1ex} & \hspace*{1ex}Energy
         (keV)\hspace*{1ex} & \hspace*{1ex}Wavelength (\AA)\hspace*{1ex} &
         \hspace*{1ex}Total photons (cm$^{-2}$ s$^{-1})$\hspace*{1ex}
         & \hspace*{1ex}Error\hspace*{1ex} \\
            \noalign{\smallskip}
            \hline\hline
            \noalign{\smallskip}

N\,{\sc vi}  &  0.430650   & 28.790 & $1.77\times10^{-5}$ & $25.50\times10^{-5}$ \\
N\,{\sc vi}  &  0.497930   & 24.900 & $1.03\times10^{-5}$ & $10.11\times10^{-5}$ \\
O\,{\sc vii} &  0.573950   & 21.602 & $3.44\times10^{-5}$ & $15.81\times10^{-5}$ \\
O\,{\sc vii} &  0.665580   & 18.628 & $7.60\times10^{-7}$ & $112.20\times10^{-7}$ \\
O\,{\sc vii} &  0.697720   & 17.770 & $1.78\times10^{-6}$ & $15.36\times10^{-6}$ \\
O\,{\sc vii} &  0.713370   & 17.380 & $1.32\times10^{-6}$ & $7.65\times10^{-6}$ \\

Ne\,{\sc ix} &  0.904990   & 13.700 & $1.31\times10^{-6}$ & $25.25\times10^{-6}$ \\
Ne\,{\sc ix} &  0.914810   & 13.553 & $2.48\times10^{-6}$ & $48.58\times10^{-6}$ \\
Ne\,{\sc ix} &  0.921950   & 13.448 & $2.10\times10^{-7}$ & $241.06\times10^{-7}$ \\
            \hline\hline
         \end{tabular}
%     $$ 

   \end{table*}
%__________________________________________________ 
%

%__________________________________________________ 
% table
   \begin{table*}
%   \centering
      \caption[]{Fit parameters of the continuum fit to the quiescent
      source spectrum (ACIS-S) (model: wabs(raymond) + 2 *
      wabs(bremsstrahlung) + 2*gauss), $\chi^2 = 3.02$. Bremsstrahlung
      temperature held fixed at value derived for the outburst spectra.} \label{Tabqcontfit}
%     $$ 
         \hspace{4cm}
         \begin{tabular}{llccc}
            \noalign{\smallskip}
            \hline
            \noalign{\smallskip}
            \hspace*{2ex}Parameter \hspace*{2ex}& \hspace*{2ex}Type of model \hspace*{2ex}& \hspace*{2ex}Unit\hspace*{2ex} & \hspace*{2ex}value \hspace*{2ex}& \hspace*{2ex}error\hspace*{2ex} \\
            \noalign{\smallskip}
            \hline\hline
            \noalign{\smallskip}

            $n_H$ &                & $10^{22}$cm$^{-2}$ & 60.6  & 10.0 \\
            kT    & bremsstrahlung & keV     &  10.56     & {\em frozen}  \\
            norm$^1$  & bremsstrahlung &     & $6.7\times10^{-4}$  & $2.8\times10^{-4}$ \\
            \hline
            $n_H$ &                & $10^{22}$cm$^{-2}$ & 1.409    & 0.070 \\
            kT    & bremsstrahlung & keV     &  10.56      & {\em frozen} \\
            norm$^1$  & bremsstrahlung &     & $2.02\times 10^{-4}$ & $0.13\times10^{-4}$ \\
            \hline
            $n_H$ &                & $10^{22}$cm$^{-2}$ & 0.000  & 0.069 \\
            kT    & raymond        & keV     & 1.072  & 0.049 \\
            norm$^2$  & raymond    &         & $6.6 \times 10^{-6}$ & $3.6\times10^{-6}$ \\
            \hline

            $E$   & Gaussian       & keV     & 0.573950 & frozen  \\       
            norm$^3$  & Gaussian   &         & $1.5\times10^{-6}$ & $0.8\times10^{-6}$\\
            \hline\hline
         \end{tabular}
%     $$ 

$^1$ norm(brems) = $3.02\times10^{-15} / (4\pi D^2) \int n_e n_I dV$,
where $D$ is the distance to the source (in cm) and $n_e$, $n_I$ are the
electron and ion densities (in cm$^{-3}$).

$^2$ $(10^{-14}/(4\pi(D_A(1+z))^2))\int n_e n_HdV$ where $D_A$ is the
angular size distance to the source (cm), $n_e$ and $n_H$ are the electron
and H densities (cm$^{-3}$).

norm(bbody) = $L_{39}/ D_{10}^2$, where $L_{39}$ is the source
luminosity in units of $10^{39}$\,erg s$^{-1}$ and $D_{10}$ is the
distance to the source in units of 10~kpc. With a distance of 460\,pc
this yields a luminosity for the black body component $L = 1.7 \times
10^{33}$~erg s$^{-1}$.

$^3$ norm (gaussian) = total photons cm$^{-2}$ s$^{-1}$ in the line.

   \end{table*}
%
%__________________________________________________ 

\section{Discussion}
\label{picture}

%-----------------------------------------------------------
   \begin{figure*}
   \centering
  \resizebox{16cm}{!}{\rotatebox{270}{\includegraphics{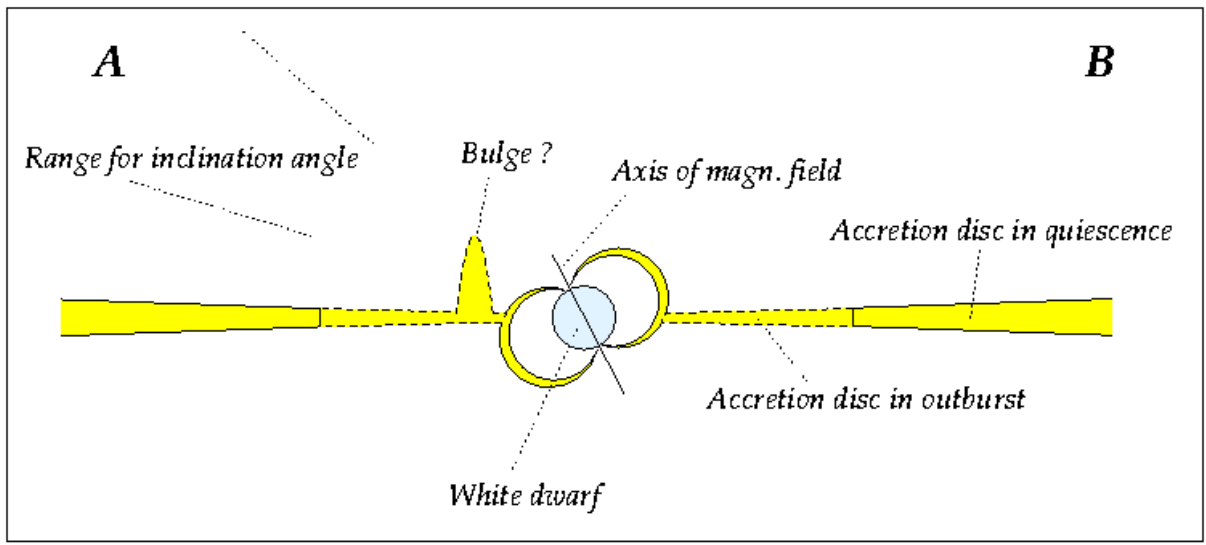}}}
      \caption{Sketch of GK~Per during outburst with cut through the white dwarf, the
      accretion curtains and the accretion disc. 
              }
         \label{Figipmodel}
   \end{figure*}
%
%______________________________________________________________

%-----------------------------------------------------------
   \begin{figure}
   \centering
    \includegraphics[width=8cm]{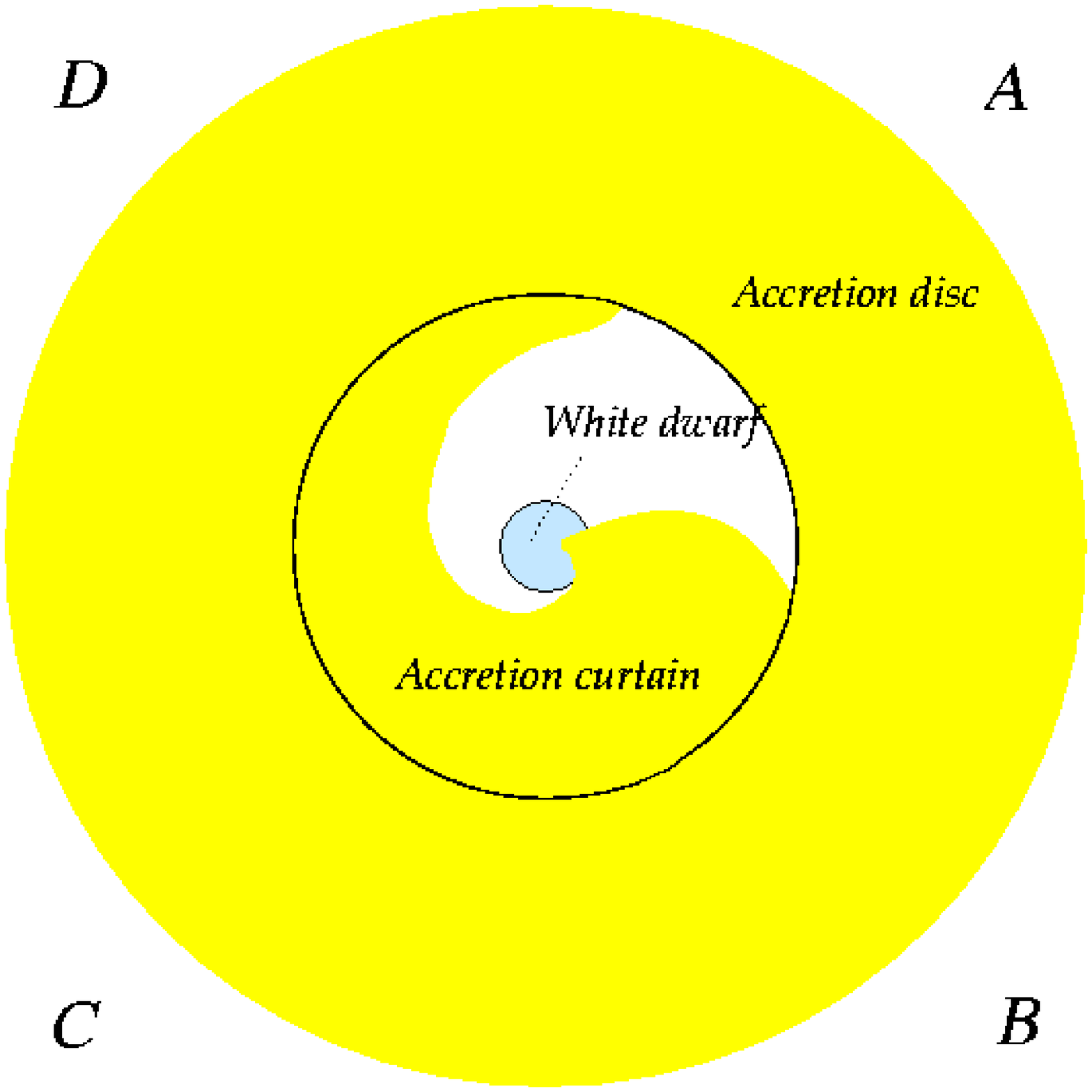}
      \caption{Sketch of the accretion disc and proposed shape of the
              accretion curtain during
              outburst in GK~Per with accretion disc seen face-on.
              }
         \label{Figcurtain}
   \end{figure}
%
%______________________________________________________________

\subsection{The model}

%model, anti-correlated QPO light curves
Fig.~\ref{Figipmodel} illustrates our model for GK~Per, similar to
that proposed by Hellier et al. (2004). During an outburst the inner
disc radius shrinks to a few white dwarf radii leading to much shorter
accretion curtains. With an inclination angle of $50^\circ$ to
$73^\circ$, the lower pole becomes obscured, reflected in the change
from a double peaked spin light curve during quiescence to a single
peaked one during outburst. The bulge rotating at the inner disc edge
does not only obscure and absorb the X-ray from the accretion region
on the white dwarf (Hellier et al. 2004), but also reprocesses the
radiation and emits it at UV wavelengths leading to maxima in UV at
times the bulge rotates into view (Fig.~\ref{Figqpolcv}). The
optical emission follows probably the UV (QPO) light curve (as Mauche \&
Robinson 2001 find correlated UV and optical DNOs for SS~Cyg).

\subsection{Asymmetry of accretion curtains}

%asymmetry of spin light curve
The variation in the X-ray spin light curve is caused by absorption
effects alone (Hellier et al.~2004), i.e., during low phase the pole
is pointing towards the observer, causing absorption of the X-ray flux
in the accretion column. Thus, a reappearance slower than
disappearance of the accretion column indicates that the curtains are
asymmetric, spread out and semi-transparent to the X-ray emission from
the foot point of the accretion column (see Fig.~\ref{Figipmodel}). The
fact that the difference in X-ray spectra during spin high and low
phase can be solely explained by a variation in the column density
supports this model.

Fig.~\ref{Figcurtain} shows a sketch of the model as seen from the
`north' pole (accretion disc face-on) and where the southern accretion
curtain is left out for clarity of the sketch. During the orbit, the
observer views the system clockwise from positions A to D. The shape
of the accretion curtain is inspired by the asymmetric spin light
curve with a sudden drop (leading bow of the accretion curtain at
phase 0.6 in Fig.~\ref{Figspinlcv}, with observer being at a position
between A and B) and slow rise (feathered out accretion curtain
wrapped nearly around the white dwarf, phase 0.9 to 1.3, position B to
D) and a short maximum (phase 0.3 to 0.5, position D to A) during
which we have an essentially undisturbed view of the accretion region
and the post-shock part of the accretion column.  The shape is caused
by the spinning white dwarf dragging material from the inner edge of
the accretion disc.  Thus, accretion takes place from nearly all
azimuths as proposed by Hellier et al. (2004).

\subsection{Blobby accretion}

%flares
The flares are interpreted as an indirect indication of blobby
accretion. A somewhat larger accumulation of material (a blob) does not
change the light curve if the total amount of accreted material is
constant in time. However, if such a large blob enters the photosphere
of the white dwarf it leads to enhanced thermal emission at this spot
compared to more homogeneously accreted material, as it will
penetrate into deeper layers causing a larger surface area of the
white dwarf to be affected.

The fact that the flares seem to occur at times when the 5000~s
(QPO) cycle shows a maximum is not coincidence if one considers that
not only the continous emission from the accretion column but also
the flared emission underlies the cyclic variation in the strength of
the absorbtion by the rotating bulge.

\subsection{The leaky absorber}

% medium resolution spectra, fit
The simple leaky absorber model describes the hard X-ray part of the
observed medium-resolution X-ray emission of the accretion region very
well both in quiescence and outburst.  The main difference between the
outburst and quiescent state is the increased absorption due to
larger column densities during quiescence as well as steeply decreased
electron $n_e$ and ion densities $n_I$ due to the enlarged curtains in
quiescence. In reality, we are likely to face a more complex physical
situation with a range of column densities or variable covering
factors (Hellier et al. 1996) due to inhomogeneous in-falling material
that would leave too many free parameters to fit for the currently
available data.
%comparison: Hellier et al. 1996, leaky absorber in AO Psc
%other systems

\subsection{Origin of emission lines}

% no variation of Fe fluorescence line on spin and QPO
The constancy of the emission lines including the Fe fluorescence line
with spin period implies that the emission site of these lines are
visible at all times and from all viewing angles during an orbit,
i.e., it ought to be in the accretion curtain. The Fe fluorescence
will also have a contribution from the white dwarf as it possibly
shows Compton scattering in the red line wing. Since we do not observe
any noticeable line shifts, the accretion curtains must be broad (due
to the assumed geometry with a spinning white dwarf and a small inner
disc radius) or accretion takes place from all azimuths because during
outburst the accretion flow overwhelms the magnetosphere as suggested
already by Hellier et al.~(2004) or due to the above mentioned
scenario.

The fact that the emission lines do also show no variation during the
5000~s (QPO) cycle (flares) implies that their emission site cannot be obscured
by the bulge. This is in accordance with the finding that the hard
X-ray component shows less variation than the soft component,
%and with the fact that the emission lines must be visible at all
%times,
i.e., they originate in the broad accretion curtains, and with the
model of accretion from all azimuths.  This again is supported by the
width of the emission lines, which are significantly broader than the
instrumental broadening (Section~\ref{emls}). The increasing width of
the lines with wavelength indicates a roughly constant velocity
dispersion for all lines, i.e., all lines originate from all parts of
the accretion curtain. However, the accretion curtains may be
structured and asymmetric as the multi-peaked emission line profiles
(especially N\,{\sc vii}) suggest.

%\subsection{Nova shell emission}
The shell emission can be fully explained by line emission from
ions of, e.g., O, N and Ne. We confirm Balman's (2001)
suggestion that the line emission is produced in the clumped ejecta in
the post-shock region, as the shell velocity of $\sim$ 1100~km
s$^{-1}$ (shell extension of 50\arcsec\ 99 years after the nova outburst at
a distance of 460~pc) results in a too high shock temperature to
produce the observed lines.

\begin{acknowledgements}
This work is based on observations obtained with {\it XMM-Newton}, an
ESA science mission with instruments and contributions directly funded
by ESA Member States and the USA (NASA), and {\it Chandra X-ray
Observatory}, operated for NASA by the Smithsonian Astrophysical
Observatory. We gratefully acknowledge the variable star observations
from the AAVSO International Database contributed by observers
worldwide and used in this research for determining at which point in
the outburst cycle observations of GK~Per were made.  S.V.
acknowledges support from DLR under 50OR0105. J.-U.N. acknowledges
support from PPARC under grant number PPA/G/S/2003/00091.

\end{acknowledgements}

\end{document}